\documentclass[preprint,epsfig,12pt]{aastex63}

\newcommand{\etal}{{\em et al.\ }}

\newcommand{\kms}{km~s$^{-1}$}

\def\ion[#1 #2]{#1\,{\sc #2}}
\def\lamb[#1]{#1\,{\AA}}
\def\lambr[#1-#2]{{{#1}--{#2}\,{\AA}}}
\def\rat[#1 #2]{#1/#2}
\def\serts89{SERTS-89}

\def\tabul{\hbox{\raise 0.75pt\hbox{$\triangleleft$}}}
\def\ergs[#1]{#1 {ergs}~{cm$^{-2}$}\,{s$^{-1}$}\,{sr$^{-1}$}}
\def\dens[#1]{10$^{#1}$\hskip 1.5pt{cm$^{-3}$}}
\def\densr[#1 #2]{10$^{#1}$\hskip 1pt{--}\hskip .5pt{10$^{#2}$}\hskip 1.5pt{cm$^{-3}$}}
\def\fl[#1 #2]{{#1}$\pm${#2}}
\def\orb[#1 #2]{{$#1^{#2}$}}
\def\ls[#1 #2]{{$^{#1}${#2}}}
\def\tm[#1 #2 #3]{{$^{#1}${#2}$_{#3}$}}

\begin{document}

\title{ACE SWICS observations of solar cycle variations of the solar wind}

\correspondingauthor{Enrico Landi}
\email{elandi@umich.edu}

\author{A. Cardenas-O'Toole, E. Landi}
\affiliation{Department of Climate and Space Sciences and Engineering, University of Michigan}

\begin{abstract}
In the present work we utilize ACE/SWICS in-situ measurements of the properties
of the solar wind outside ICMEs in order to determine whether, and to what 
extent are the solar wind properties affected by the solar cycle. We focus on
proton temperatures and densities, ion temperatures and differential speeds, 
charge state distributions and both relative and absolute elemental abundances. 
We carry out this work dividing the wind in velocity bins to investigate how 
winds at different speeds react to the solar cycle. We also repeat this study, 
when possible, to the subset of SWICS measurements less affected by Coulomb 
collisions. We find that with the only exception of differential speeds (for 
which we do not have enough measurements) all wind properties change as a function
of the solar cycle. Our results point towards a scenario where both the slow 
and fast solar wind are accelerated by waves, but originate from different 
sources (open/closed magnetic structures for the fast/slow wind, respectively) 
whose relative contribution changes along the solar cycle. We also find that the 
signatures of heating and acceleration on one side, and of the FIP effect on 
the other, indicate that wave-based plasma heating, acceleration and fractionation 
remain active throughout the solar cycle, but decrease their effectiveness in 
all winds, although the slow wind is much affected than the fast one. 
\end{abstract}


\section{Introduction}

The solar wind consists of a continuous stream of supersonic particles flowing 
from the Sun into the Heliosphere, which shapes the physical properties of the 
interplanetary space and plays a fundamental role in Space Weather. In fact, 
the solar wind sets the stage for the propagation of Interplanetary Coronal 
Mass Ejections (ICMEs), influences some of their properties and their arrival 
time at the Earth, and in part it determines the overall effects that ICMEs 
have on the Earth's magnetic field as Space Weather events. Also, the solar 
wind is capable of generating Space Weather events of its own, when faster 
wind streams overtake slower ones and create streams of shocked materials 
which can also affect the Earth's magnetic field. Space Weather events 
represent a serious threat to our technological society, which is only destined 
to grow as modern civilization becomes more and more reliant on energy distribution,
communication systems and space assets that can be directly damaged or even
disabled by the effects of a Space Weather event such as ground induced currents,
energetic particles, and ionospheric disturbances. The importance of the solar 
wind makes it imperative to develop a Space Weather forecast system capable 
of giving advance warning of an incoming solar storm and a robust 
understanding of the solar wind phenomenon itself.

The solar wind is a phenomenon most likely tied to the magnetic field in 
the atmosphere of the Sun. In fact, different magnetic field configurations
are thought to give rise to different types of wind (e.g. Cranmer 2009 and
references therein). Still, the solar magnetic field undergoes an 11-year
cycle that radically changes its morphology and in turn the properties of
the solar wind it sustains. The most spectacular indication of such a time 
dependent variation comes from the measurements carried out by the Ulysses 
spacecraft during its three polar passes in 1996 (solar minimum), 2001 
(solar maximum) and 2008 (solar minimum) clearly showing how the spatial 
distribution of fast ($v > 700$~\kms) and slow ($v < 500$~/ms) wind 
completely changes during solar maximum and minimum (McComas \etal 1998,
2008). As in-situ measurements of the solar wind date back for decades, 
multiple studies have helped determine its properties evolve during the 
solar cycle.

\section{Solar cycle dependence of solar wind properties}

\subsection{Magnetic field}

After a certain distance from the photosphere, the solar wind plasma 
$\beta$ is larger than 1, so that the wind plasma drags the solar magnetic 
field along as it travels in the heliosphere. This magnetic field is very 
important as it interacts with planetary magnetospheres during both 
quiescence and storms. Since the magnetic field of the Sun changes along 
the solar cycle, the question is whether also the magnetic field of the 
solar wind shows a solar cycle variability. Zerbo \& Richardson (2015) 
studied 27-day running averages of the solar wind magnetic field strength
as measured near Earth and compared them with the variations of the sunspot 
number (as a proxy of the solar cycle) for the last five solar cycles (20 
to 24). They found that the wind average magnetic field strength does 
correlate with the phase of the solar cycle, being larger during maximum 
and smaller during minimum. Furthermore, they found that the distribution 
of daily averages measured during a 1-year interval was the same at each 
of the solar minima with the only exception of the minimum of cycle~24, 
when the shape of the distribution was still the same but peaked at a 
lower value (3~G instead of 5~G). On the contrary, the distributions 
during solar maxima were very similar.

Owens \etal (2016a) utilized indirect proxies of the solar wind magnetic field
to extend the period over which its time variation could be studied, reaching as 
far back as 1750. They provided estimates of the annual average wind magnetic
field and successfully compared their results with direct estimates from the 
OMNI spacecraft observations. They confirmed the correlation between wind field
magnitude and sunspot number found by Zerbo \& Richardson and showed that a 
similar correlation existed as far down as 1750. Furthermore, they also showed 
that longer term variations exist, where the wind magnetic field minimum and 
maximum values change over multiple cycles, showing that the very low value
of cycle~24 is actually similar to the values obtained around 1900, and that the
large maximum values from 1950 to 1990 (with the exception of 1970) are similar
to those occurred in 1860. Unfortunately, the 250-year time span they covered 
is still too short to determine whether multiple periodicities were present. 
In all cases, the wind magnetic field values varied in the 4-10~G range, showing 
that the solar cycle can alter the average wind field by a factor at least 2.

\subsection{Dynamical properties}

Zerbo \& Richardson (2015) also tried to correlate the 27-day averages of the
wind speed, density and mass flux with the sunspot number, finding that
the correlation was much weaker than for the magnetic field and on the overall
not much significant. Using the same dataset, Yermolaev \etal (2021) found
an overall decrease of the main plasma parameters in the solar wind from
cycle 22 to 23, which have then remained low during cycles 23 and 24. Still, 
a marked solar cycle distribution of the source regions of the wind near the
ecliptic was noted by Luhmann \etal (2002), who found that polar coronal holes 
contributed to the wind in the ecliptic for only half of the solar cycle around 
solar minima, while during maxima the wind came from equatorial sources.

However, the OMNI measurements they used only allowed for
measurements near the Earth. Using a completely different method, Lamy \etal
(2017) were able to paint a much more complete picture for the mass flux. They
used Ly-$\alpha$ backscattering measured by the SOHO/SWAN instrument (Bertaux
\etal 1995) from 1996 to 2014 to determine the mass flux of the solar wind. 
Lamy \etal (2017) obtained maps of the distribution of the solar mass flux
with longitude for every Carrington Rotation from 1996 to 2014. They found
that the mass flux had a clear heliolatitudinal distribution, which changed 
as a function of the solar cycle phase: during minimum, the largest portion 
of the mass flux was concentrated within 20 degrees from the equator, with 
the fast wind filling the rest of the heliosphere with a much smaller mass 
flux; during maximum the slower, denser solar wind increased the absolute 
value of the mass flux, and was distributed over all latitudes. 

The solar cycle dependence of both the latitudinal coverage and the absolute 
value of the mass flux cause the total solar mass loss to the solar wind to 
strongly depend on the solar cycle. An independent estimate of the total 
solar mass loss rate by Wang (1998) indicated that it varied from 
2$\times 10^{-14}~M_{sun}/yr$ at the minimum to 3$\times 10^{-14}~M_{sun}/yr$ 
at the maximum during solar cycles 21 and 22. The variations in magnitude 
and spatial distribution of mass loss to the solar wind is very important, 
as the presence of denser solar wind material at all latitudes during solar 
maximum causes the rate of cosmic rates reaching the Earth to decrease, 
while during minimum the decreased protection of solar wind material causes 
the rate of galactic cosmic ray arrival at Earth to increase. For example, 
Fludra (2015) showed that the number of galactic cosmic ray hits affecting 
the detectors of the SoHO/CDS instrument orbiting at L1 was clearly dependent 
on the solar cycle, with the very weak minimum of cycle 24 recording a larger 
number of cosmic ray hits than the previous, stronger minimum.

Last, solar cycle variations in wind density and distribution also affect
the torque exerted by the solar wind on the Sun. Finley \etal (2018) utilized
three different methods to calculate the angular momentum loss of the Sun, 
yielding results that, while providing different average values, all clearly
indicated a marked solar cycle variation, with the value of the torque 
increasing by a factor 3-5 from minimum to maximum.

\subsection{Elemental composition}

Elemental abundances are a very important solar wind parameter for two main reasons. 
First, once the solar wind accelerates away from the Sun, they are supposed to remain 
unaltered as the plasma travels into the Heliosphere, so that the values measured 
in-situ can be used to identify the plasma parcel's source region. Second, elemental 
abundances in the lower solar atmosphere are different from their photospheric 
counterparts: differences depend on each element's first ionization potential (FIP) 
being larger (high-FIP) or smaller (low-FIP) than 10~eV.  In fact, the low-FIP/high-FIP 
ratio measured in the solar corona is enhanced by a factor (called {\em FIP bias}) 
dependent on the region's magnetic field configuration. Usually, in open field 
configurations such as coronal holes the FIP bias is around unity (that is, no 
enhancement) while in closed field configurations it is anywhere between 2 and 5, 
possibly depending on time. These abundance changes, called FIP effect (Laming 2015), 
are thought to be due to ponderomotive forces tied to the same type of magnetic waves 
that are also one of the candidate entities responsible for solar wind heating and 
acceleration. Thus, studying the elemental abundances of the solar wind can provide 
precious information also on magnetic waves at the base of the solar corona, where 
the elemental fractionation takes place. 

There are two ways of measuring solar composition: the abundance of an element 
relative to Hydrogen, also known as {\em absolute abundances} (e.g. C/H), and 
the abundance of an element relative to another element other than Hydrogen, 
named {\em relative abundances}, e.g. Ne/O. The former are tied to the 
metallicity of the Sun, the latter are affected by the FIP effect.

Since the FIP effect seems to be related to the magnetic field, it is fair to expect
that it might be affected by the solar magnetic cycle. Several authors investigated 
the variation of abundance ratios of several key elements. For example, Kasper \etal
(2007, 2012) utilized WIND measurements of the $\alpha$/proton ratio as a function
of time from 1995 to 2010, thus capturing two solar minima and the maximum of solar 
cycle~23, finding a very strong dependence. Most importantly, they found that the 
time variation of this ratio was strongly dependent on the wind speed: the slower 
the speed, the larger the solar cycle sensitivity. With the only exception of the 
wind faster than 560~\kms, which showed a remarkable stability throughout the solar 
cycle, the $\alpha$/proton ratio decreased during minimum and reached a maximum of 
around 0.05 during maximum: the slowest speed wind ($v<300$~\kms) decreased this 
ratio by a factor 5-10 and the ratio was lower during the minimum of solar cycle~24 
in 2008 than it was in the minimum of cycle~23 in 1996. To make the 2008 minimum 
even more peculiar, even the fastest wind showed a decrease in the $\alpha$/proton 
ratio during 2008-2009 that it did not show during the previous minimum of 1996-1997. 
The analysis of Kasper \etal (2008,2012) stops at 2010, but results were extended 
until 2015 and confirmed for the rising phase of cycle~24 by Zerbo \& Richardson 
(2015).

Lepri \etal (2013) carried out a similar study utilizing data from the ACE/SWICS
instrument in the 1998-2011 range, to determine the solar cycle variation of the 
abundances of heavy elements detected in-situ at 1~AU. They only divided the solar 
wind in two velocity classes (faster and slower than 500~\kms) and found that there 
is a $\approx$50\% decrease in the abundances of He, C, O, Si, and Fe relative to H 
as the Sun moved towards the minimum of cycle~24, indicating a decrease of the wind's 
metallicity; also, the FIP bias in the fast solar wind showed some variations 
indicating changes in the efficacy of the FIP effect in the fast wind's source 
regions. Other signs of solar-cycle induced changes in the wind composition were 
found by Shearer \etal (2014) studying the Ne/O abundance ratio: this ratio is of 
critical importance for FIP fractionation models, for the Ne and O photospheric 
absolute abundances, and for helioseismological models of the solar interior. 
Shearer \etal (2014) found that the Ne/O ratio changed as a function of the solar 
cycle by factors also dependent on wind speed, being almost constant in the fast 
wind, and changing by a factor $\approx$1.5 in the slow wind: a similar variability 
was found by Landi \& Testa (2015) in remote sensing observations of quiescent 
streamers, a possible source for the slow wind.

\subsection{Charge state distributions}

In-situ measurements of the charge state distribution of the solar wind have long 
been utilized to infer the conditions of the solar coronal region where the wind
comes from. In fact, the solar wind undergoes electron-ion collisional ionization 
and recombination as it is accelerated from the source region in the heliosphere, 
traveling through a very steep initial temperature gradient in the solar transition 
region. As the plasma density decreases with distance, the efficiency of collisional 
processes decreases until a point is reached when the electron density is sufficiently 
low to effectively stop collisional ionization and recombination from occurring. From 
this point onward, the ionization status of the solar wind does not evolve and remains 
the same throughout the heliosphere. This point is called "freeze-in" point, and its 
physical location depends on each element and each ion, as the ionization and 
recombination rates are different for species. Once frozen in, 
the charge state distribution of solar wind elements maintains a record of the 
electron density and temperature (which determine the local ionization and 
recombination rates) it has experienced before reaching the freeze-in point, 
as well of the wind speed, which determines how much time the wind spends in 
the densest regions where it can be ionized/recombined.

The most comprehensive study of the solar cycle dependence of the solar wind charge
state distribution has been carried out by Lepri \etal (2013), who studied the charge
state ratios of C, O and the average charge state of Fe. They found that both the
fast and slow solar wind change their ionic composition as a function of the solar
cycle, decreasing it during solar minimum; changes were largest for the less dense 
fast wind, but significant for both. Differences were large enough to affect the
use of charge state ratios such as $O^{7+}/O^{6+}$ to discriminate between fast
and slow solar wind, as slow wind ratio values at solar minimum were the same as 
fast wind values during solar maximum. Landi \& Lepri (2015) investigated
the effect of solar-cycle variations of photoionizing X-ray, EUV and UV flux on the
solar wind ionization status, finding that it accounted only for part of the
variation, the rest being likely due to a change electron in density, temperature 
and wind speed.

\subsection{Goal of the present work}

The results obtained by Kasper \etal (2007, 2012) and Lepri \etal (2013) clearly 
demonstrate that the solar wind compositional plasma properties do depend on the 
solar cycle. However, the former focused on $\alpha$ particles only, while Lepri 
\etal (2013) did not consider the two quantities that can provide information on 
the wind heating and acceleration processes: ion temperatures and differential 
velocities. Also, these studies did not investigate how collisional age (e.g. 
Tracy \etal 2015, 2016), an indication of the amount of Coulomb collisions 
experienced by a wind plasma parcel, affect the measured solar wind properties.
Furthermore, Lepri \etal (2013) divided the solar wind in two broad velocity 
classes (separated at 500~\kms); however, as the dependence of ion temperatures 
and ionization status on wind speed is continuous rather than a step function, 
such broad velocity bins lose resolution in determining the speed dependence of 
these quantities.

Our goal is to improve on these earlier works by:

\begin{enumerate}

\item Dividing the wind in much finer velocity classes, to increase velocity resolution;
\item Including ion temperatures and differential wind speed in the analysis;
\item Investigating how the correlation of ion temperatures with charge-to-mass ratios, 
mass, and wind speed depends on the solar cycle;
\item Determining how lower collisional age influences the results.

\end{enumerate}

\noindent
The data and methodology we use are described in Section~\ref{data}; results are reported 
in Section~\ref{results} and discussed in Section~\ref{discussion}, while
 Section~\ref{conclusions} summarizes this work.

\section{Data and methodology}
\label{data}

%
%
%
%
%
%
%
%
%

\subsection{Instruments}

The data used in the present work primarily come from the SWICS instrument (Gloeckler
\etal 1998) on board the ACE satellite. We utilized the data collected from the start 
of the mission in 1998 to 2011, after which an anomaly in the hardware increased the 
background and generated several invalid measurements. The 1998-2011 time interval 
covers one full solar cycle, from the rising phase of cycle~23 to the rising phase 
of cycle~24. The data we used consist of bulk velocity, thermal speed, and density 
for the most abundant ions observed by SWICS: He$^{2+}$, C$^{4-6+}$, N$^{5-7+}$, 
O$^{5-7+}$, Ne$^{6-9+}$ (with the exception of Ne$^{7+}$, unavailable), and 
Fe$^{7-12+}$; we have used 2-hr averaged data. The thermal speed in this dataset 
only includes the component along the SWICS look direction.

Proton data have also been rebinned from the original 12-minute resolution to 2~hours 
by averaging them and re-normalizing in case some time stamps were missing; however, 
care was taken to ensure that the proton plasma properties were constant during each 
2-hour time stamp. We first removed data where the error was larger than 0.4, then we 
removed all time stamps for which either the proton speed or proton temperature averages 
had a standard deviation larger than half of each quantity's value.

Differential speeds between protons and heavy ions were measured only from plasma streams
traveling within 10$^o$ of the magnetic field direction. To select this sample, we utilized
magnetic field measurements from the MAG instrument (Smith \etal 1998) on board ACE. Given 
the large variability of the magnetic field within two hours, we only kept the time stamps 
where the magnetic field was relatively constant; we selected these by calculating average 
magnetic field $B_{tot}$ and the standard deviation $\sigma_B$ and requiring that 
$sigma_B/B_{tot}<0.3$: this additional requirement greatly decreased the number of time 
stamps available for differential speed studies.

%

\subsection{Data selection}

To improve the quality of the sample, for each ion we utilized only time stamps with a number 
of counts larger than 15, except to determine the total abundance of an element, in which 
case we accepted time stamps where at least one ion of this element had more than 15 counts.
Due to the fact that Oxygen is usually clustered in the O$^{6+}$ charge state (which usually
accounts for at least 80\% of the element in all wind types) we did not use Oxygen in element
abundance studies when O$^{6+}$ did not have the required 15 counts.

Since we are interested in the background solar wind only, we have removed data taken during
ICME events using the Richardson \& Cane list (Cane \& Richardson 2003, Richardson \& Cane 2010); 
we further excluded wind streams not present in that list, which were either faster than 800~km/s, 
or with Fe$^{16+}$/Fe ratio values larger than 0.1 or O$^{7+}$/O$^{6+}$ ratio values larger than 1. 


In order to study the dependence of wind's properties on the solar cycle with a high
velocity resolution, we divided the 300-650~\kms velocity range in 14 velocity bins,
each 25~\kms wide; we further created two more bins for the fast wind (650-700~\kms
and 700-800~\kms) and one bin collecting all wind slower than 300~\kms. The width of 
these bins have been selected as a compromise between achieving high velocity resolution, 
while at the same time maintaining a significant number of events in each velocity bin.
However, due to the lack of data in some velocity bins (e.g. 325-350~\kms in 2003 or 
650-700~\kms in 2009), some gaps in the dataset are present.

\subsection{Methodology}

We study several properties of the solar wind as observed close to the ecliptic plane 
by ACE/SWICS in the entire 1998-2011 range. We aim at characterizing the behavior of 
an array of plasma properties with the goal of understanding whether and how the solar 
cycle affects the heating, acceleration, and evolution of the solar wind. 

The element composition of the wind was studied both to determine the wind's absolute 
and relative abundances. We study the latter to understand whether the abundance ratio 
of low-FIP to high-FIP elements (that is, the {\em FIP bias}) changes with time to 
investigate the FIP effect (Laming 2015). Also, we seek to understand whether the wind 
acceleration mechanisms are as effective at extracting heavy ions into the solar wind 
during the solar cycle as they are at accelerating protons.

Charge state composition provides information on the main quantities that determine 
the ionization status of the solar wind: plasma electron density, electron temperature 
and speed. The former two quantities determine the local efficency of free electrons 
at ionizing wind ions at any location before the freeze-in distance, while the latter 
determines the time that the wind spends being ionized at each location (Landi \etal 
2012).

Ion temperatures and differential velocities are two signatures for wave-based wind 
heating and acceleration (Cranmer \etal 2008, Khabibrakhmanov \& Mullan 1994, Cranmer 
\etal 1999, Chandran 2010, Kasper \etal 2013). In particular, ion kinetic temperature 
ratios to proton temperatures are expected to depend on ion parameters such as mass
and charge/mass ratio due to the effect of waves, while heavy elements are predicted 
to stream faster than protons by up to an Alfven speed. Solar cycle dependence of 
these two parameters would indicate effects on the efficiency of waves at accelerating 
and heating the solar wind. However, these signatures tend to be erased by Coulomb 
collisions, which work toward equalizing ion and proton speeds, as well as ion and 
proton temperatures. For this reason, we have repeated, where the number of available 
data points made it possible, the study of these two quantities limiting the sample 
to plasma parcels whose {\em collisional age} $A_C$ (Tracy \etal 2015, 2016) is less 
than 0.3, where $A_C$ is defined as:

\begin{eqnarray}
A_C & = & 2.19\times 10^7\frac{Z_i^2}{A_i}\frac{\ln\Lambda_{ip}}{i{\left({\frac{T_i}{A_i}+T_p}\right)}^{3/2}}\frac{n_p}{v_p} \\
\log\Lambda_{ip} & = & 29.9-\ln{{\left[{\frac{Z_i{\left({A_i+1}\right)}}{A_iT_p+T_i}{\sqrt{\frac{n_iZ_i^2}{T_i}+\frac{n_p}{T_p}}}}\right]}}
\end{eqnarray}

\noindent
where $Z_i$ and $A_i$ and the ion's charge and atomic number, $T_i$ and $T_p$ are the
ion and proton temperatures (in K), $n_i$ and $n_p$ are the ion and proton densities 
(in cm$^{-3}$) and $v_p$ is the proton speed (in~\kms). Large values of $A_C$ indicate
that the wind plasma has undergone many Coulomb collisions before reaching the instrument,
small values indicate a more pristine solar wind.

\section{Results}
\label{results}

\subsection{Proton properties}

Both proton temperature and density change as a function of time along the solar 
cycle. Also, the type and amount of change are different in different velocity 
classes.

\begin{figure}
\centerline{\includegraphics[height=10.0cm,width=14.0cm]{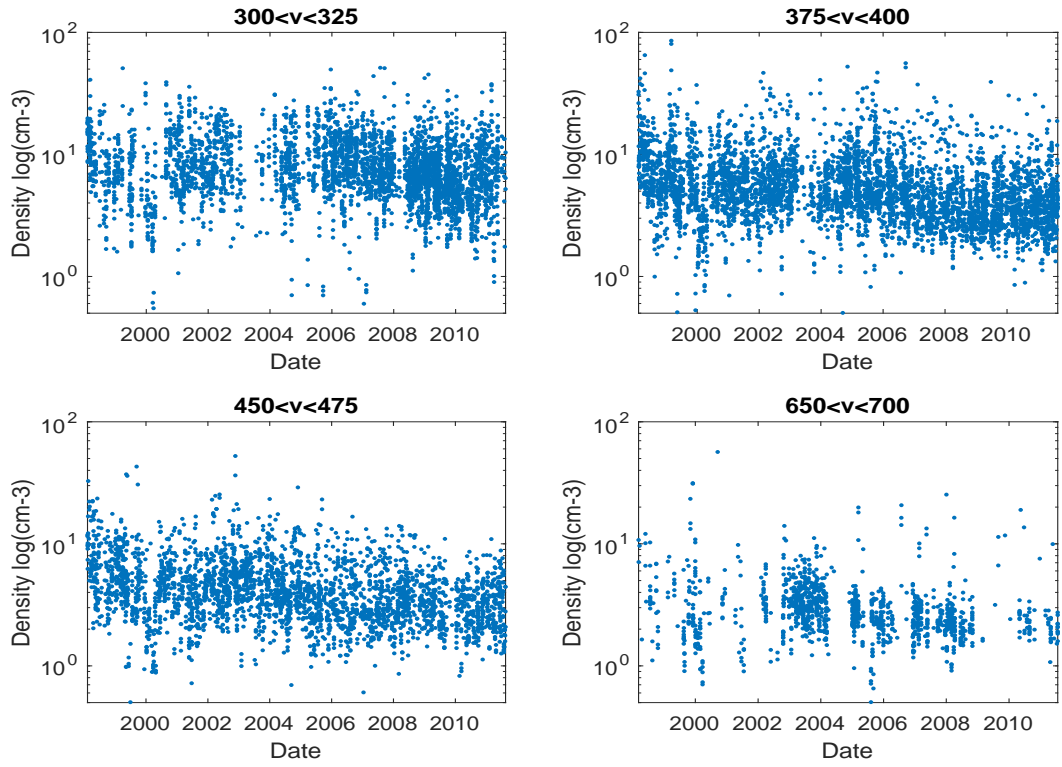}}
\caption{Proton density versus time for 4 velocity classes: slow wind (top row),
intermediate wind (bottom left) and fast wind (bottom right).
\label{proton_density}}
\end{figure}

Proton density shows a decrease over time which has two specific properties, as 
shown in Figure~\ref{proton_density}. First, the decrease is most evident at speeds
larger than 350~\kms, while at lower speeds the proton density seems to be essentially
constant. Second, the decrease is steady from 1998 to 2011, with no apparent correlation
with the solar cycle phase, so that it seems to occur at time scales much larger than 
the cycle itself. Most remarkably, it does not show any hint of recovery after the
minimum of cycle 24 in 2008. 

\begin{figure}
\includegraphics[height=8.0cm,width=11.0cm]{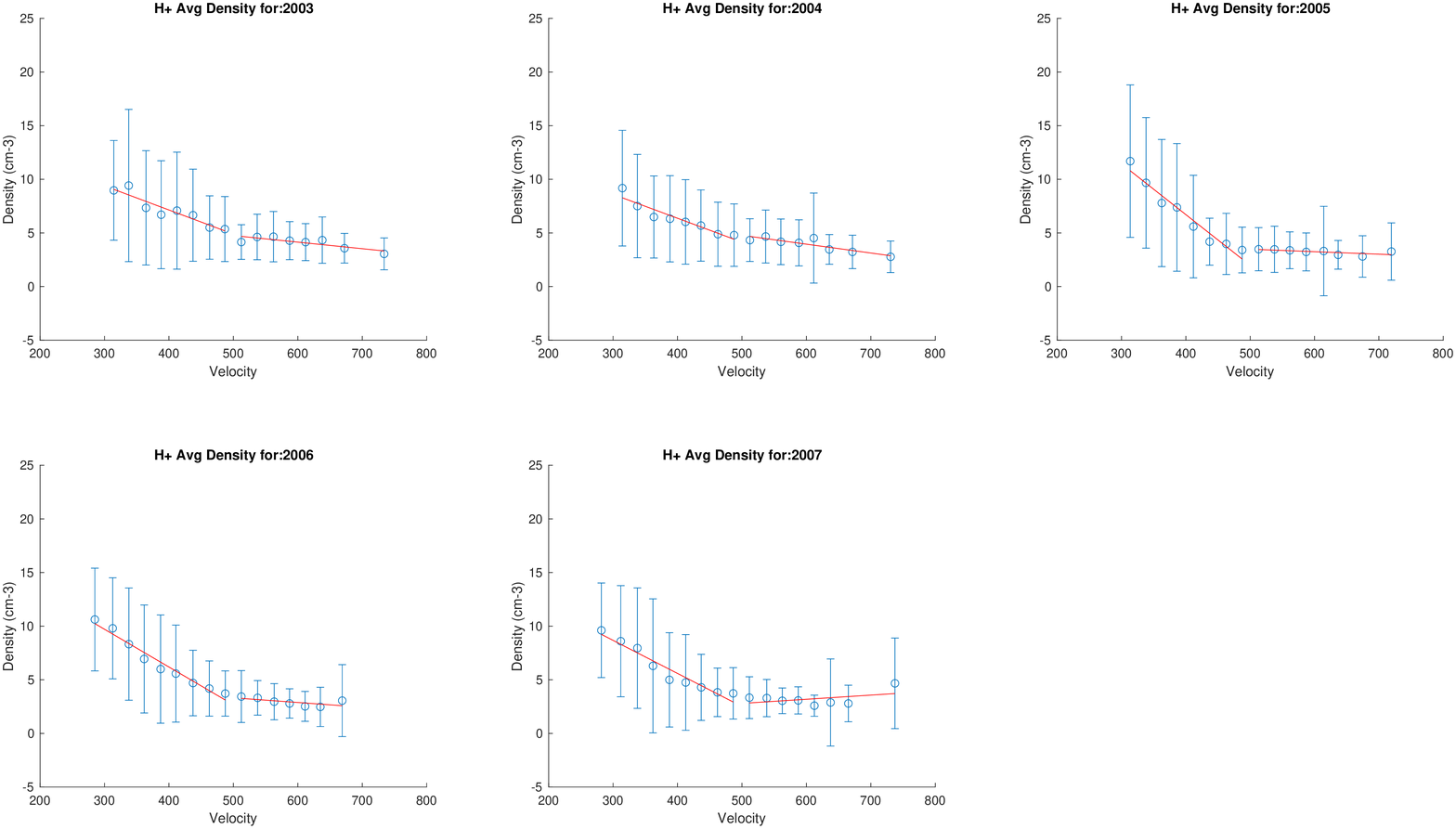}
\includegraphics[height=8.0cm,width=6.0cm]{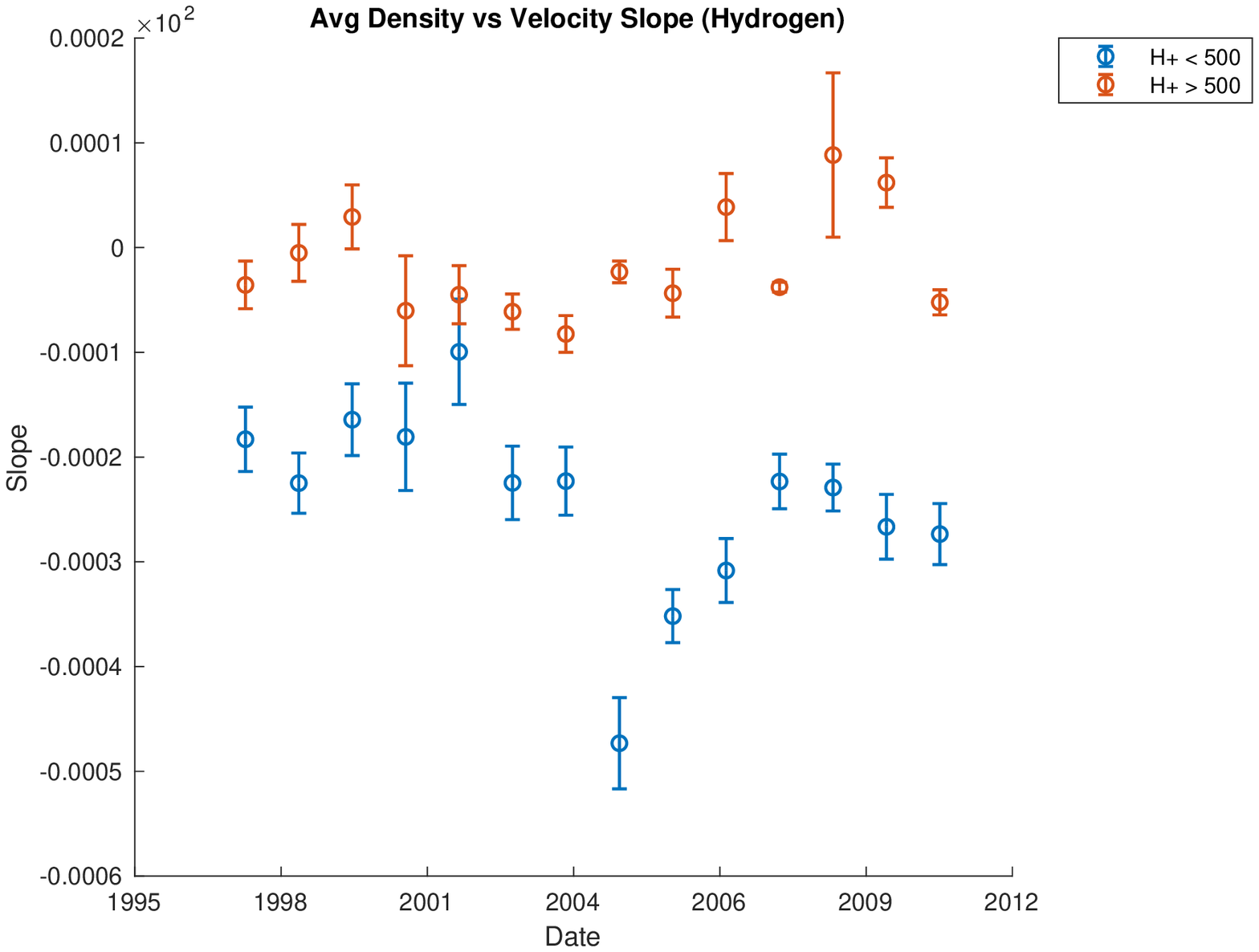}
\caption{{\bf Left:} Examples of double linear fit to the yearly averaged proton
density vs. speed, for the years 2003 to 2007. {\bf Right:} Slope of the linear 
fit to the yearly average proton density versus year. Red: velocity bins faster 
than 500~\kms. Blue: velocity bins slower than 500~\kms. See text for details.
\label{proton_density_fit}}
\end{figure}

Proton density has been long known to decrease as the wind speed increases: faster 
winds are more tenuous. We tested whether the solar cycle had any influence on 
the rate of such decrease: for each year, we averaged the proton density in each 
velocity bin and calculated a linear fit between average density and speed. We 
found that no satisfactory fit could be found that reproduced the whole range of 
speeds, as the density versus speed relationship seems to have a marked change in 
slope at around 500~\kms, as shown in Figure~\ref{proton_density_fit} (left): the 
proton density decreases faster below 500~\kms, while it seems to be almost constant 
at larger speeds. We then carried out two separate linear fits at speeds larger and 
smaller than 500~\kms, and show the slope of both fits as a function of time in the 
solar cycle in Figure~\ref{proton_density_fit} (right). Fitted slopes are different, 
with the wind faster than 500~\kms having an approximately constant density, but
neither shows a significant dependence on the solar cycle, being approximately 
constant throughout the entire period under consideration. 

\begin{figure}
\centerline{\includegraphics[height=10.0cm,width=14.0cm]{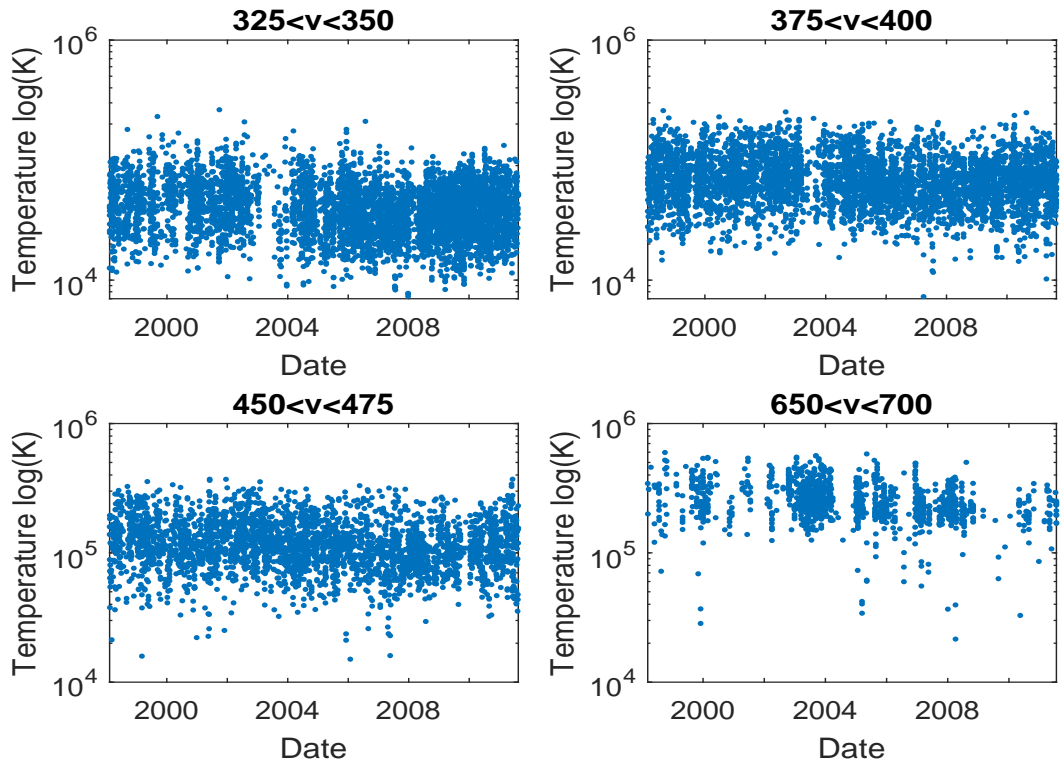}}
\caption{Proton temperature versus time for 4 velocity classes: slow wind (top row),
intermediate wind (bottom left) and fast wind (bottom right).
\label{proton_temperature}}
\end{figure}

\begin{figure}
\centerline{\includegraphics[height=12.0cm,width=18.0cm]{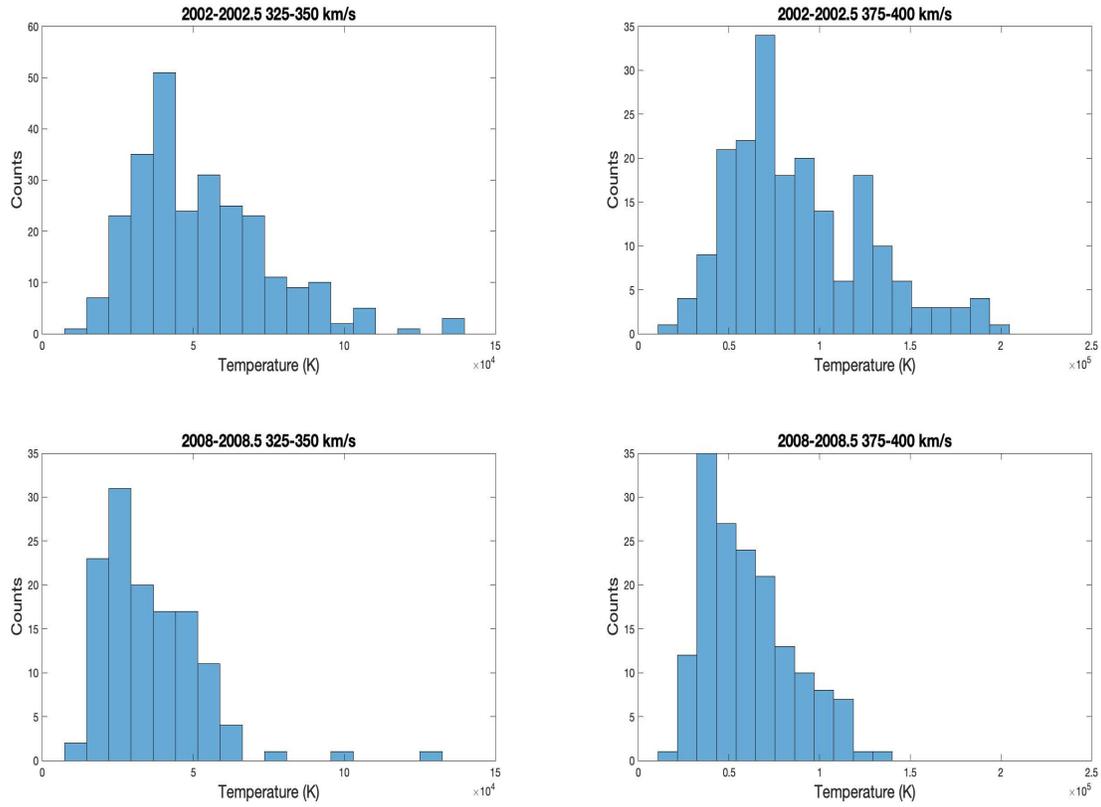}}
\caption{Histogram of proton temperatures for the 325-350~\kms (left column) and
375-400~\kms (right column) velocity classes, during solar maximum (year 2002-2002.5,
top row) and solar minimum (year 2008-2008.5, bottom row).
\label{proton_histograms}}
\end{figure}

Proton temperatures ($T_p$), on the contrary, are essentially constant with time at 
speeds larger than 400~\kms, with a value increasing with speed itself, as shown in 
Figure~\ref{proton_temperature}. On the contrary, at slower speeds $T_p$ shows some 
dependence on the cycle phase, being lower during the 2008-2009 minimum and larger 
during the 2000-2002 maximum. An example is shown in Figure~\ref{proton_histograms}, 
where at solar maximum $T_p$ peaks at larger temperatures (a bit less than a factor 2) 
and has a broader distribution than at solar minimum. At larger speeds the difference 
decreases until it disappears.

\subsection{Elemental abundances}

\begin{figure}
\centerline{\includegraphics[height=10.0cm,width=16.0cm]{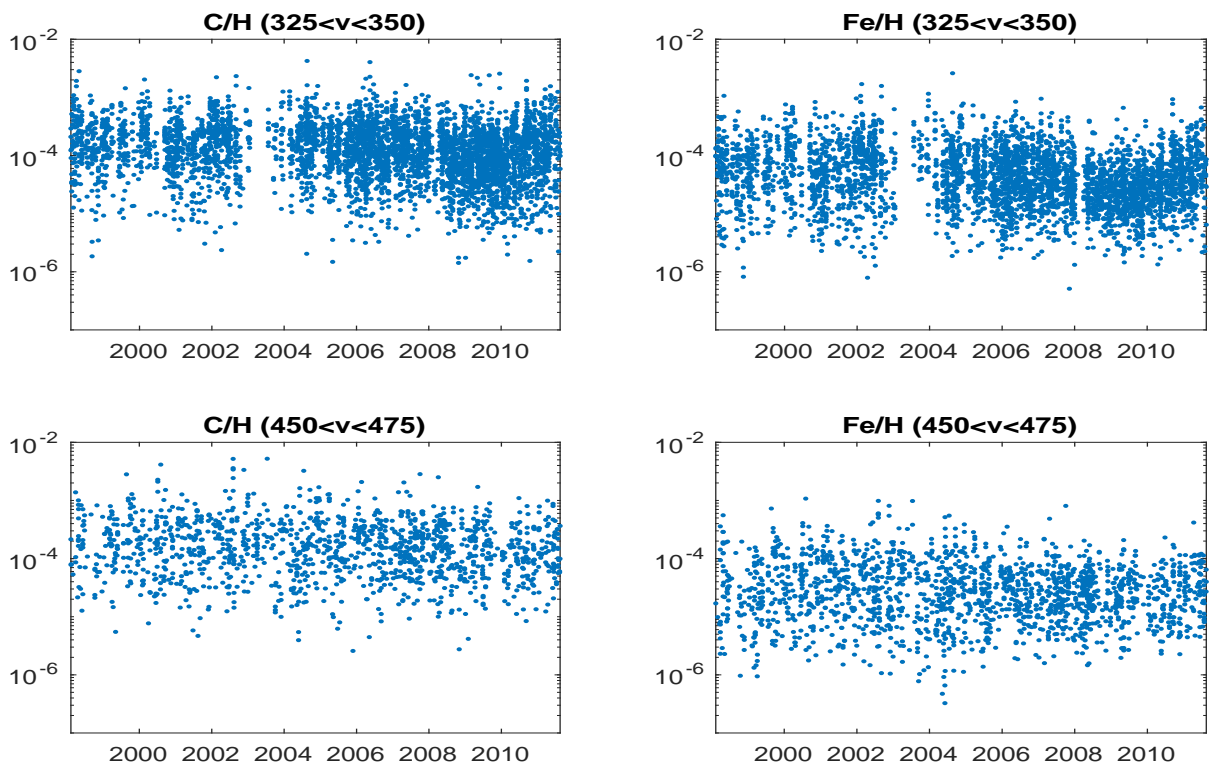}}
\caption{C/H (left panels) and Fe/H (right panels) abundance ratios as a function 
of time for two velocity classes: slow wind (325-350~\kms) and intermediate wind 
(450-475~\kms).  \label{absolute_abundances}}
\end{figure}

As far as absolute abundances are concerned, solar cycle variability occurs at 
speeds lower than 450~\kms, with the magnitude of change decreasing as the wind 
speed increases, as shown in Figure~\ref{absolute_abundances}. In the slowest velocity 
bins, the abundances of C, N, O, Ne and Fe decrease around 2009, and seem to have 
recovered by 2010. At the lowest speeds, the decrease amounts to a factor 2-3. 

The relative C/O abundance ratio is constant at $\approx$0.7 along the solar cycle, 
and has the same value in all velocity classes. Considering that the photospheric 
value of this ratio ranges between 0.45 and 0.55, this measurement indicates a FIP 
enhancement of C over O of $\approx$1.3-1.6; the stability of this value over the 
solar cycle indicates that the process enhancing C over to O is not affected 
by the solar cycle and is the same in all wind source regions. 

\begin{figure}
\centerline{\includegraphics[height=10.0cm,width=16.0cm]{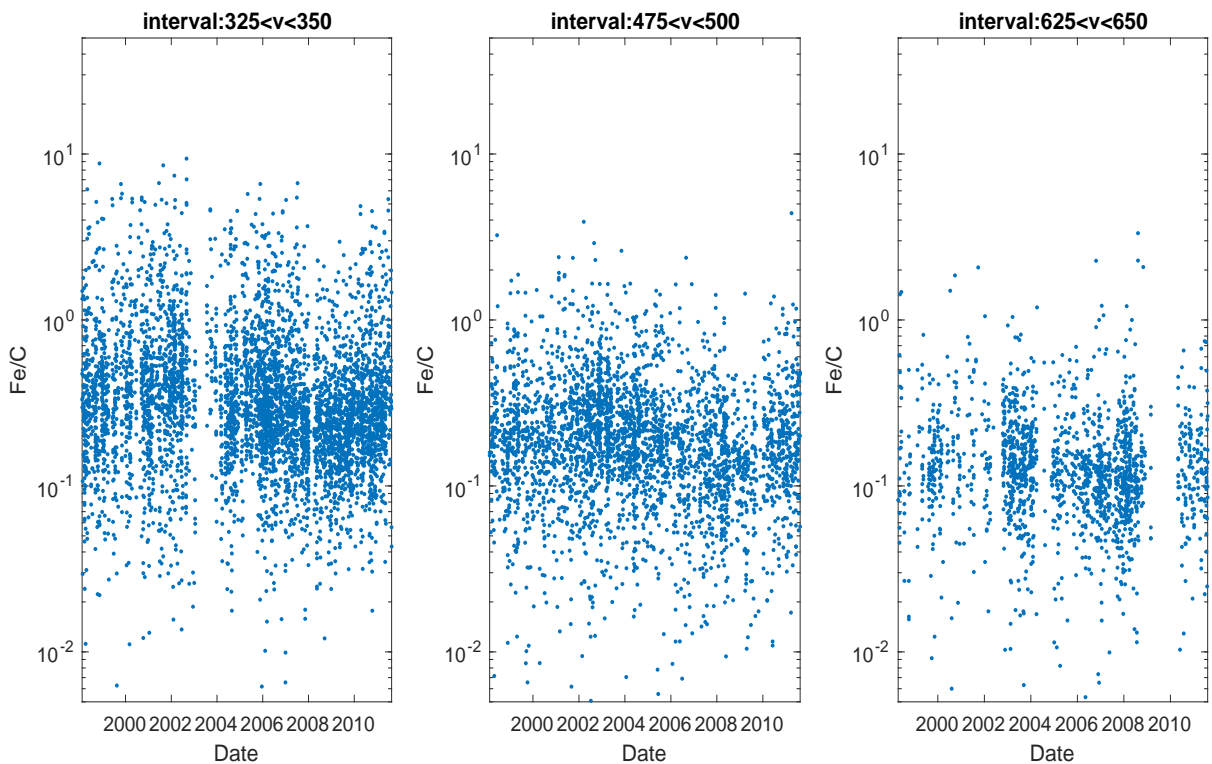}}
\caption{Relative Fe/C abundance as a function of time for slow (325-350~\kms, left),
intermediate (475-500~\kms, middle) and fast (625-650~\kms, bottom) wind.  
\label{relative_abundances}}
\end{figure}

\begin{figure}
\centerline{\includegraphics[height=10.0cm,width=18.0cm]{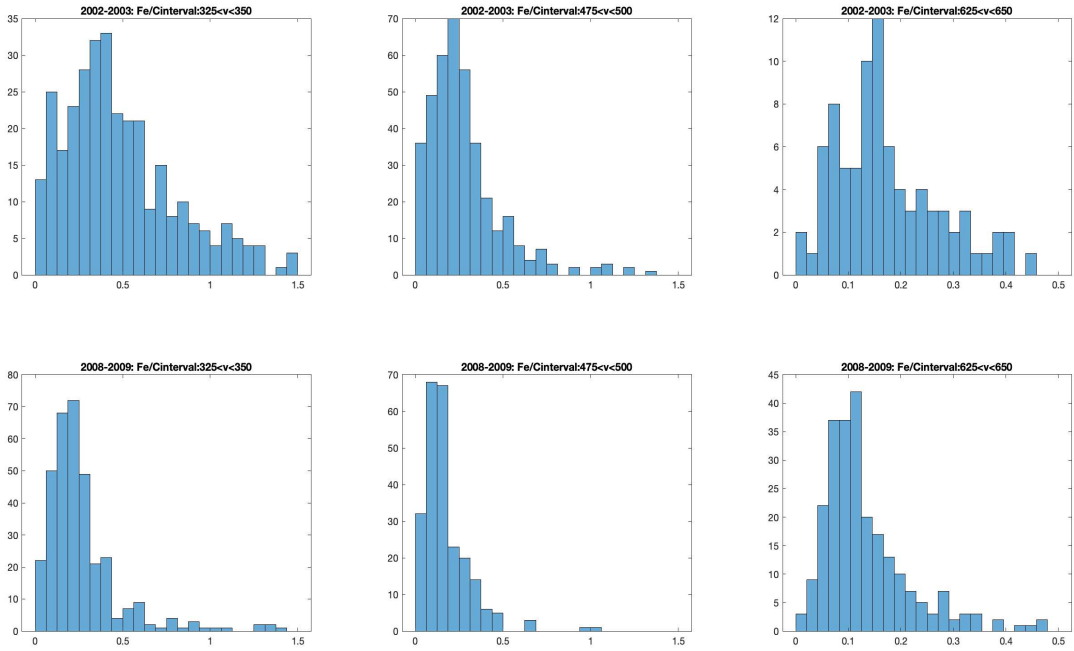}}
\caption{Histograms of the Fe/C abundance ratio during solar maximum (2003, top row)
and minimum (2008, bottom row), for the 325-350~\kms (left column), 475-500 (middle 
column) and 625-650~\kms (right column) velocity bins.
\label{abundance_ratio_histogram}}
\end{figure}

The Fe/C relative abundance, on the contrary, changes as a function of time, as 
shown in Figure~\ref{relative_abundances}. Fe/C ratio values are lower during 
solar minimum, as noted by Lepri \etal (2013). Figure~\ref{abundance_ratio_histogram} 
shows Fe/C abundance ratio histograms for three different velocity bins during 2003 
and 2008, providing a more quantitative estimate of the change; the effect is lower 
at larger speeds. It is also worth noting that the spread of values in these ratios 
is larger at lower speed, especially during solar maximum, and it decreases at larger 
speed. This is consistent with possible multiple sources of the slow solar wind, 
originating from plasmas with different composition (e.g. Stakhiv \etal 2015, 2016 
and references therein).

\begin{figure}
\centerline{\includegraphics[height=9.0cm,width=18.0cm]{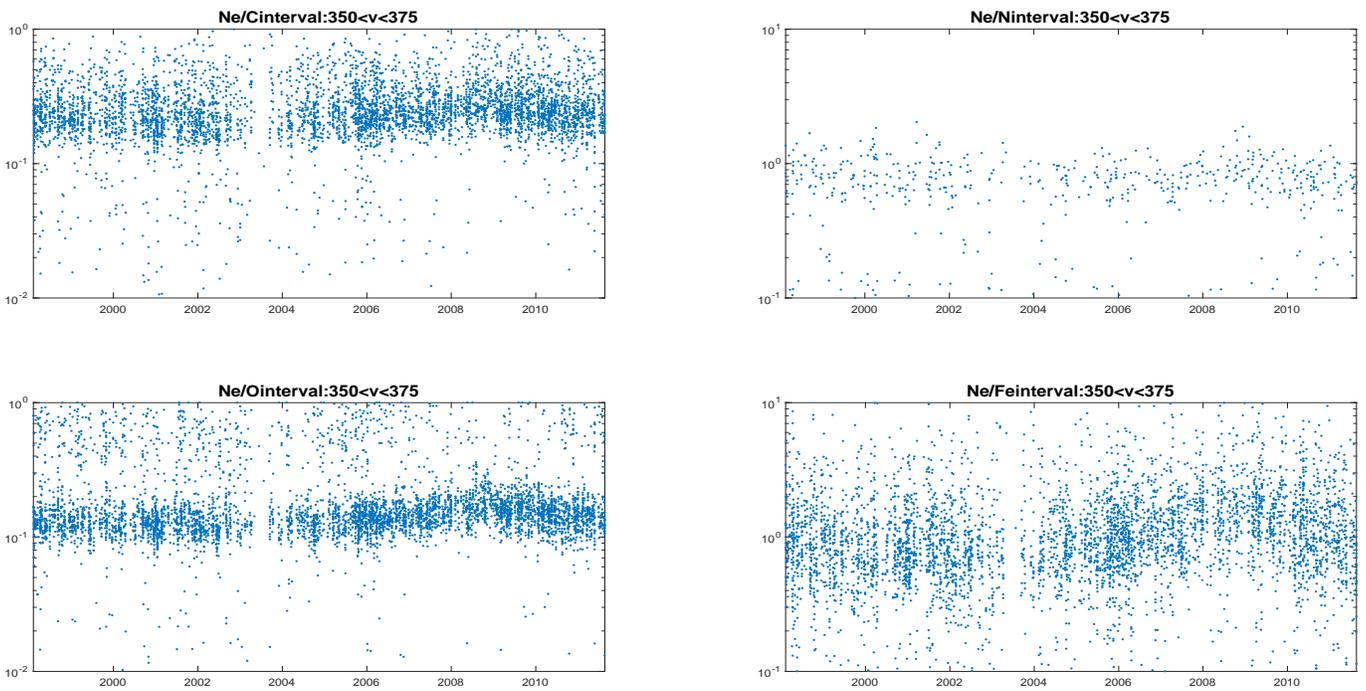}}
\caption{Ne abundance ratios as a function of the solar cycle for the 350-375~\kms
velocity bin: Ne/C (top left), Ne/N (top right), Ne/O (bottom left) and Ne/Fe 
(bottom right).
\label{ne_ratios}}
\end{figure}

\begin{figure}
\centerline{\includegraphics[height=9.0cm,width=18.0cm]{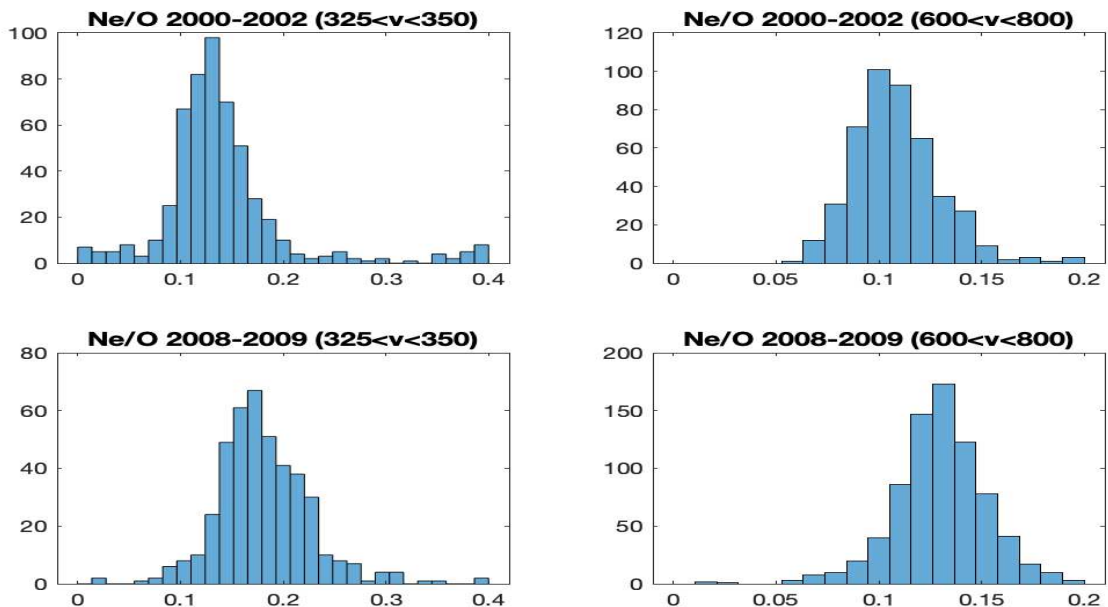}}
\caption{Histograms of the Ne/O abundance ratio during solar maximum (2000-2002, top row)
and minimum (2008-2009, bottom row), for the 325-350~\kms (left column) and 650-700~\kms 
(right column) velocity bins.
\label{ne_histograms}}
\end{figure}

Neon is a peculiar element because the Ne$^{+7}$ charge state is hopelessly blended
with the far more abundant O$^{6+}$ charge state, to the point that it is not possible
to determine its distribution from the raw measurements and thus measure its density. 
Shearer \etal (2014) discuss that the contribution of this charge state to the total 
Ne abundance is likely lower than 15\% at solar minimum, where its abundance is expected 
to be largest, and less at solar maximum. Furthermore, Neon's second most abundant 
isotope ($^{22}$Ne) accounts for $\approx$6.5\% of the total Ne abundance, and this 
value seems to undergo fractionation between the fast and the slow solar wind (Heber 
\etal 2012). Thus, the elemental abundance of Ne is subject to additional uncertainties 
than other elements. With this in mind, the abundance ratios involving Ne have a 
peculiar behavior. The relative Ne/O abundance ratio is remarkably constant throughout 
the solar cycle, despite a small increase during the minimum of cycle 23 during 
2008-2009, as shown in Figure~\ref{ne_ratios}; the increase is dependent on speed, 
being largest at a factor around 2 at the slowest speeds and decreasing at speeds 
of 500~km/s and larger (e.g. Figure~\ref{ne_histograms}). This trend was 
first described by Shearer \etal (2014), and resembled the Ne/O abundance ratio in 
quiescent streamers measured by Landi \& Testa (2015). Interestingly, the Ne/C and 
Ne/N abundance ratios show a smaller increase at solar minimum than Ne/O. The Ne/Fe 
ratio instead combines both the apparent increase of Neon over the other high-FIP 
elements, and the decrease of Fe over those same elements, and thus shows a significant 
increase at low speeds during the 2007-2009 minimum. No discernible trends were found 
for the Ne/H ratio, as its value ranges over one order of magnitude at any speeds, 
and this variability masks any possible trend with the solar cycle.

\subsection{Ion temperatures and temperature ratios}

\subsubsection{Absolute values}

\begin{figure}
\includegraphics[height=10.0cm,width=16.0cm]{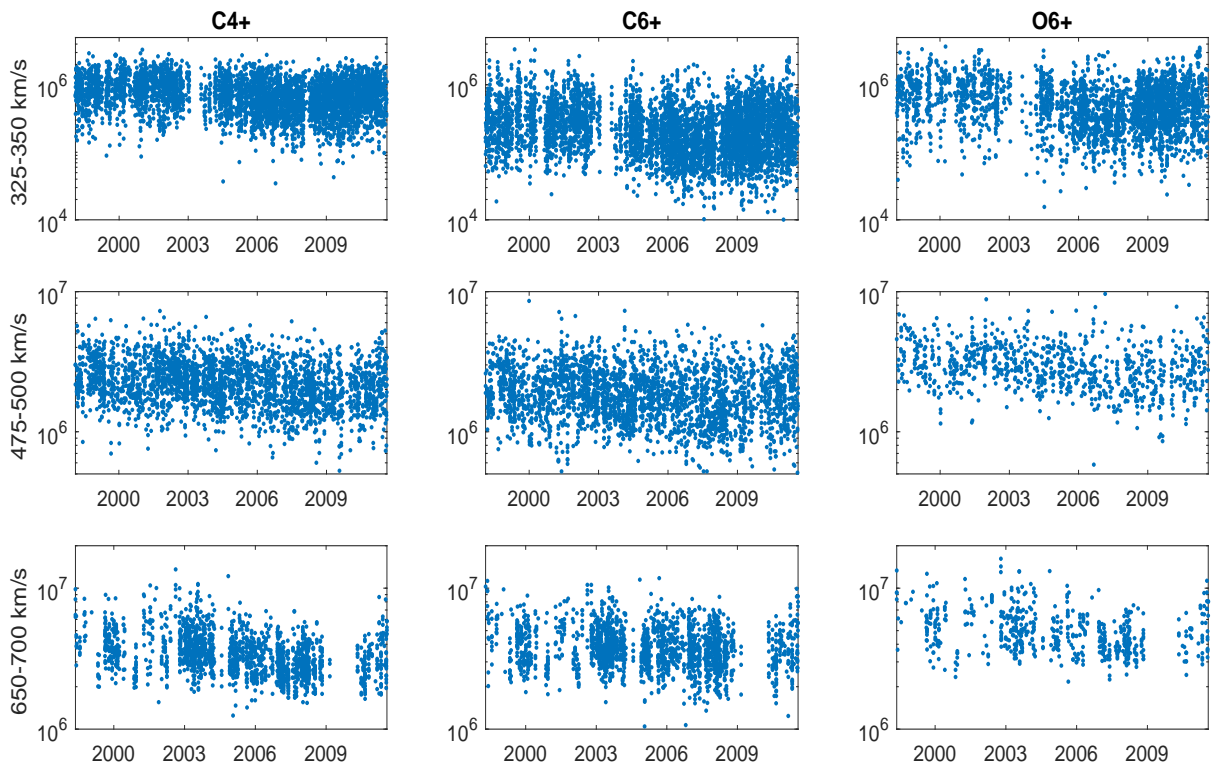}
\caption{Ion temperature values for select C and O ions as a function of time for three 
velocity classes: slow (325-350~\kms), intermediate (450-475~\kms) and fast (650-700~\kms) 
wind.
\label{tion_1}}
\end{figure}

\begin{figure}
\includegraphics[height=10.0cm,width=16.0cm]{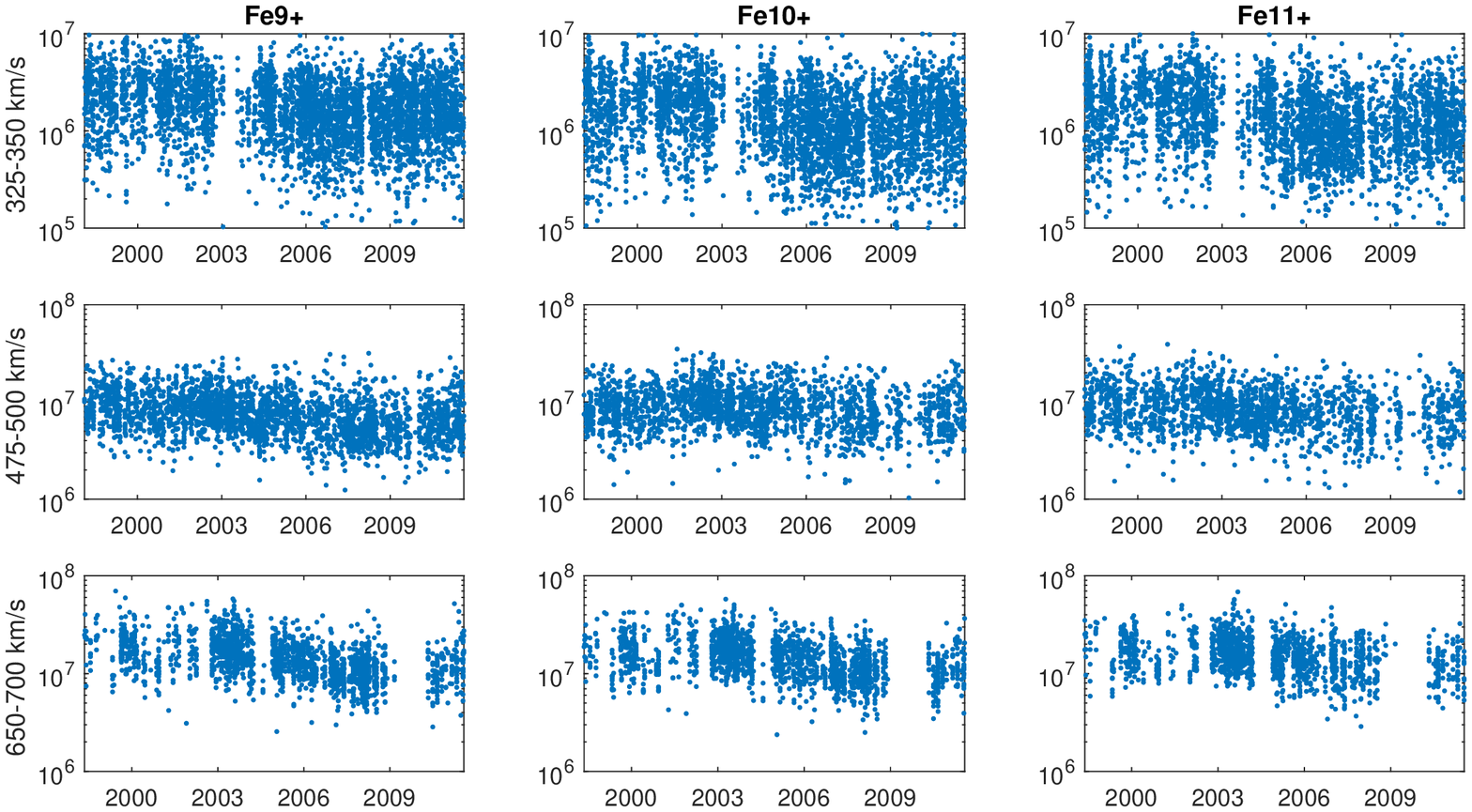}
\caption{Ion temperature values for select Fe ions as a function of time for three 
velocity classes: slow (325-350~\kms), intermediate (450-475~\kms) and fast (650-700~\kms) 
wind.
\label{tion_2}}
\end{figure}

The temperatures $T_i$ of all ions show at least some degree of variability with the 
solar cycle. Some examples are shown in Figures~\ref{tion_1} and \ref{tion_2}, which 
report values in three velocity classes (325-350, 475-500 and 650-700~\kms) for a few 
Carbon, Oxygen and Iron ions. Ion temperatures show a factor $\approx$1-3 decrease 
during solar minimum at all speeds, although the lack of solar wind in the faster 
speed bins between 2009 and 2010 prevents a definitive conclusion. Unfortunately, the 
lack of data before 1998 makes it impossible to determine whether such decrease occurs 
also during the previous minimum or was peculiar to the unusually weak cycle~24 minimum.

\begin{figure}
\includegraphics[height=10.0cm,width=18.0cm]{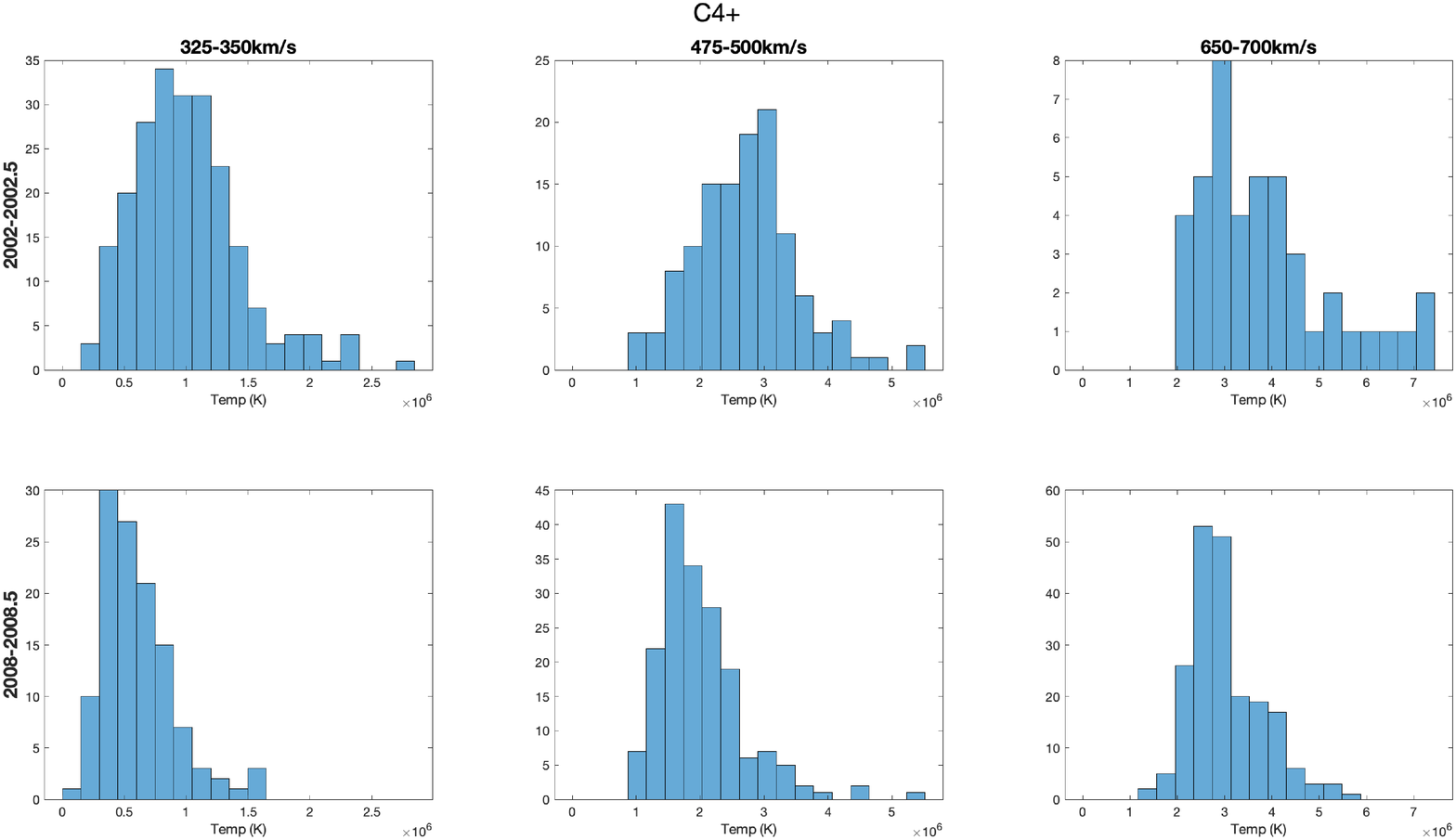}
\includegraphics[height=10.0cm,width=18.0cm]{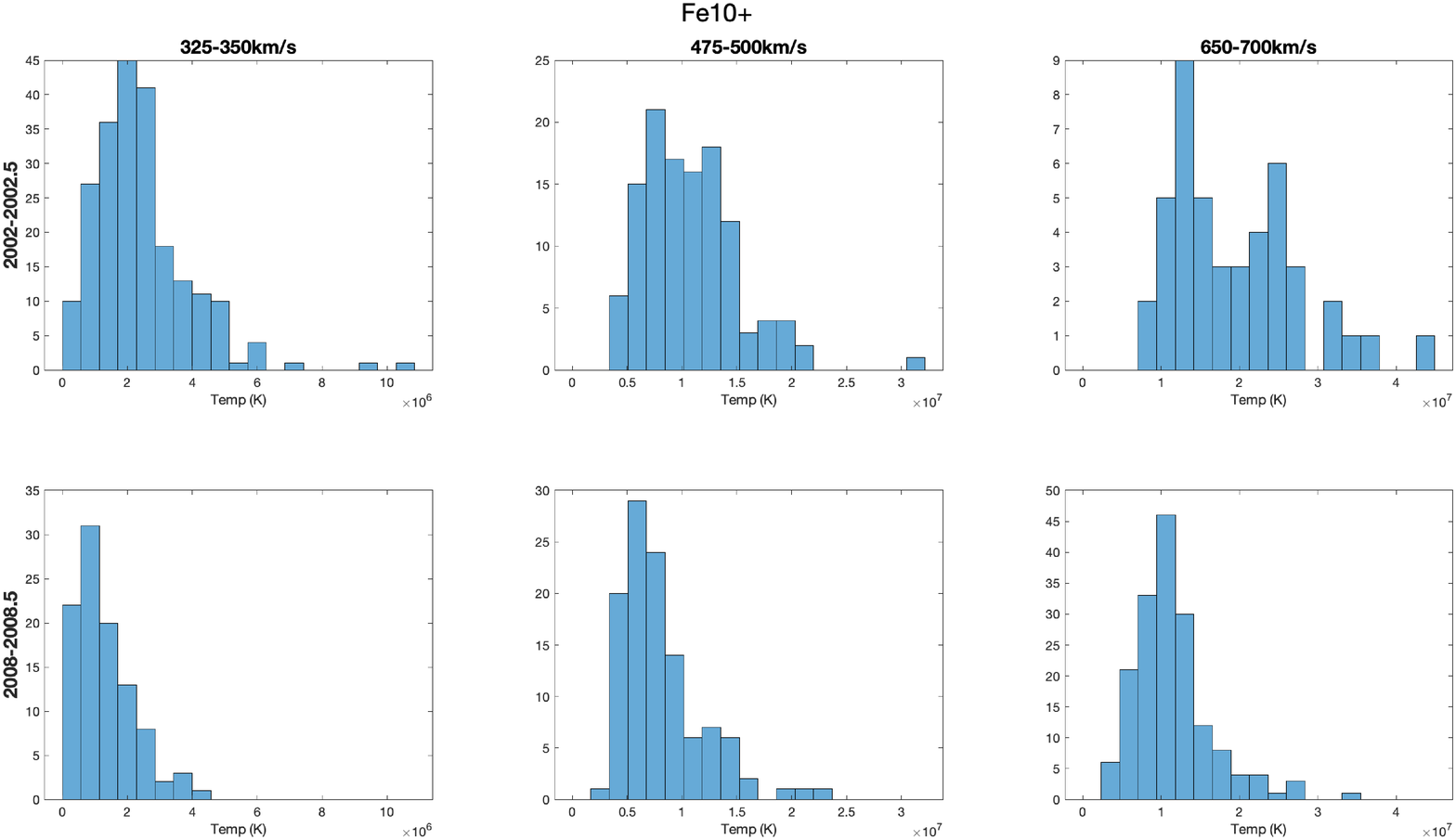}
\caption{Histograms for the C$^{4+}$ and Fe$^{10+}$ ion temperature values in 2002-2005 
(first and third rows) and 2008-2008.5 (second and bottom row) for three velocity classes: 
slow (325-350~\kms), intermediate (450-475~\kms) and fast (650-700~\kms) wind.
\label{tion_histograms}}
\end{figure}

Figure~\ref{tion_2} also shows that the spread in temperature values for each of the 
Fe ions displayed is much larger at lower speed, with values spanning approximately 
one order of magnitude while at larger speeds covering approximately a factor of 
three. Such a behavior is present to a lesser extent also in C and O, as shown in 
Figure~\ref{tion_histograms} for C$^{4+}$ and Fe$^{10+}$. 

\subsubsection{Ion temperature vs speed}

It has long been known that ion temperatures are larger in the fast wind, and decrease
along with the wind speed -- Figures~\ref{tion_1} to \ref{tion_histograms}
report some examples. The question is whether this empirical relationship changes 
along the solar cycle. To investigate this, for each ion we have averaged $T_i$ in 
yearly bins for every velocity class, and fitted the average T$_{ion}$ vs speed 
relationship every year. We experimented with fitting two types of linear functions:

\begin{eqnarray}
T_{ion} & = & a_{lin}\times v + b_{lin} \label{linfit} \\
\log T_{ion} & = & a_{log}\times v + b_{log} \label{logfit}
\end{eqnarray}

\begin{figure}
\centerline{\includegraphics[height=8.0cm,width=16.0cm]{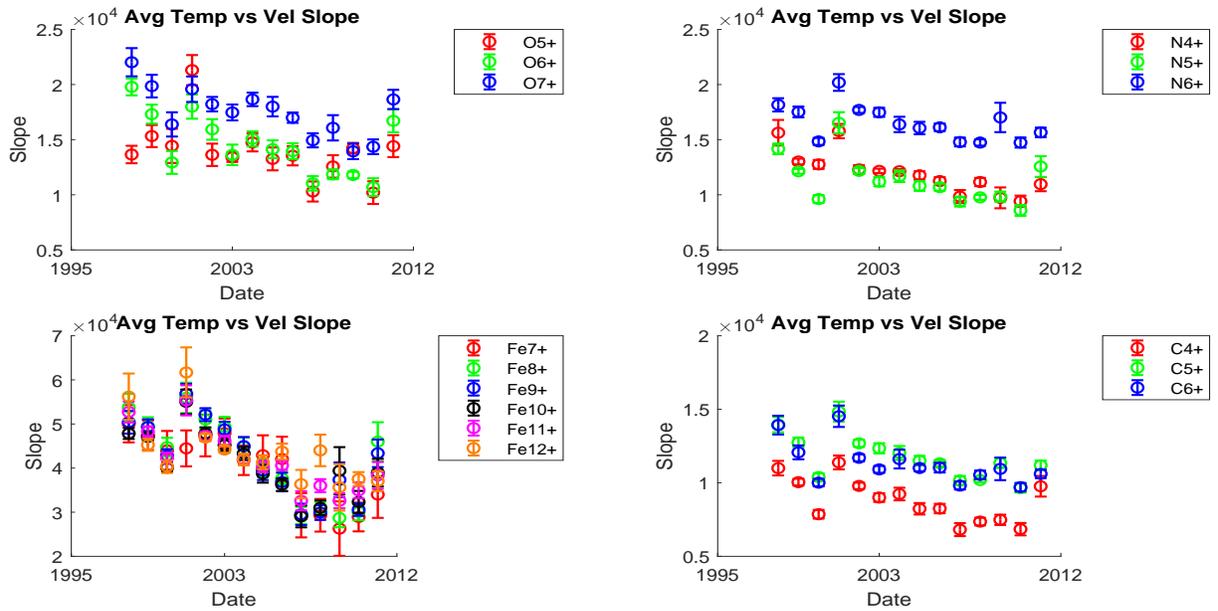}}
\caption{Value of $a_{lin}$ for C, N, O and Fe as a function time along the solar cycle.
\label{tion_vs_vel}}
\end{figure}

\begin{figure}
\includegraphics[height=8.0cm,width=8.0cm]{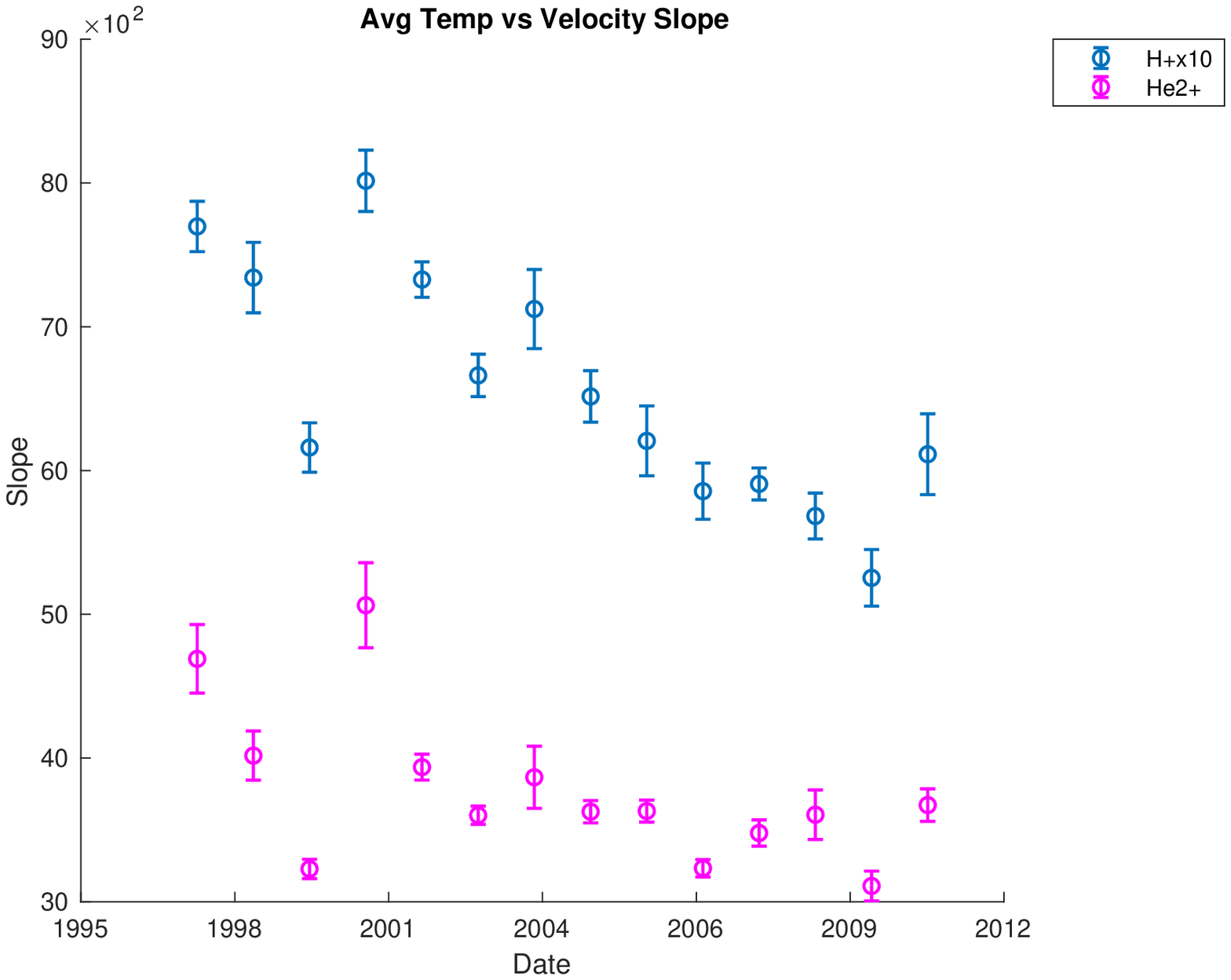}
\hspace{0.5cm}
\includegraphics[height=8.0cm,width=9.0cm]{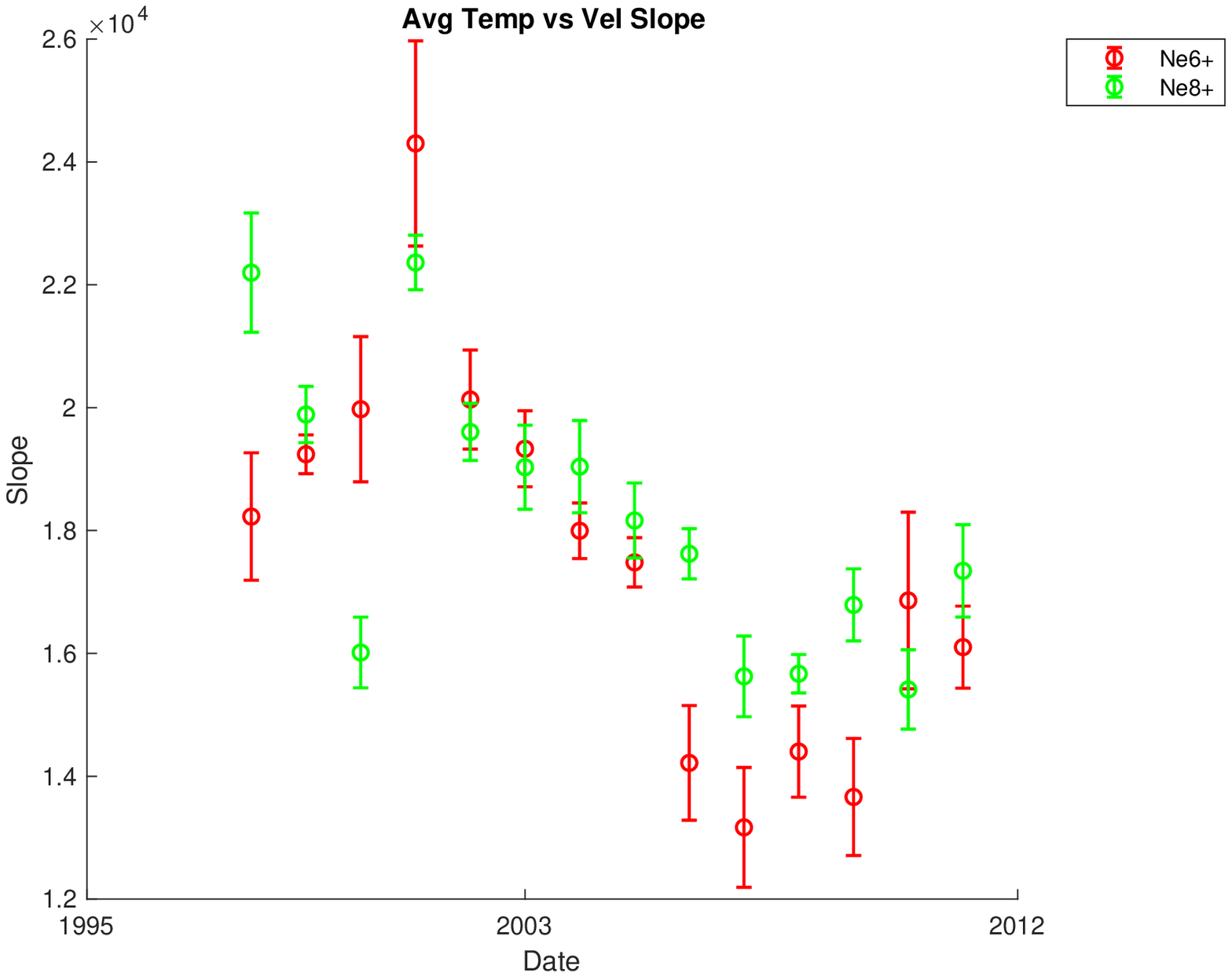}
\caption{Value of $a_{lin}$ for protons (left) and select Neon ions (right) as a 
function time along the solar cycle.  \label{tp_vs_vel}}
\end{figure}

\noindent 
We found that the correlation coefficient for Equation~\ref{linfit} was higher, so 
we analyzed the dependence of $a_{lin}$ along the solar cycle, which is shown in 
Figure~\ref{tion_vs_vel} for C, N, O and Fe, and Figure~\ref{tp_vs_vel} for H, He
and Ne. Both figures indicate that there is a tendency of $a_{lin}$ to decrease 
from cycle~23 maximum to minimum in all these elements, the only exception being
He$^{2+}$ whose $a_{lin}$ value is essentially constant after 2002. Although the 
SWICS dataset we used only includes a small part of the beginning of solar cycle~24, 
there is some hint of a recovery of the pre-minimum slope values after 2009.
This result is consistent with the ion temperature decreasing during solar minimum.


\subsubsection{Ion temperature vs charge-to-mass ratio}

\begin{figure}
\includegraphics[height=8.0cm,width=16.0cm]{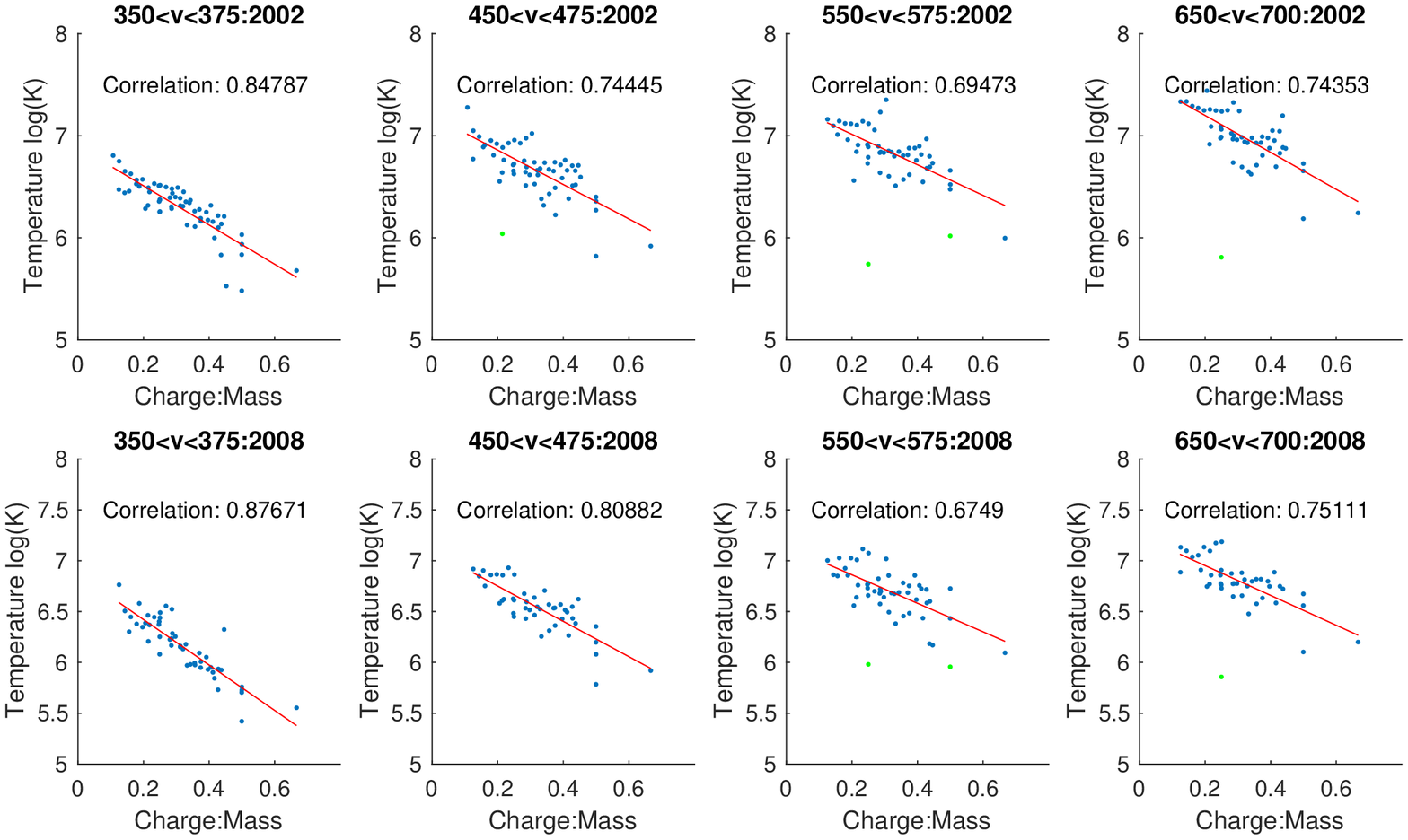}
\includegraphics[height=8.0cm,width=16.0cm]{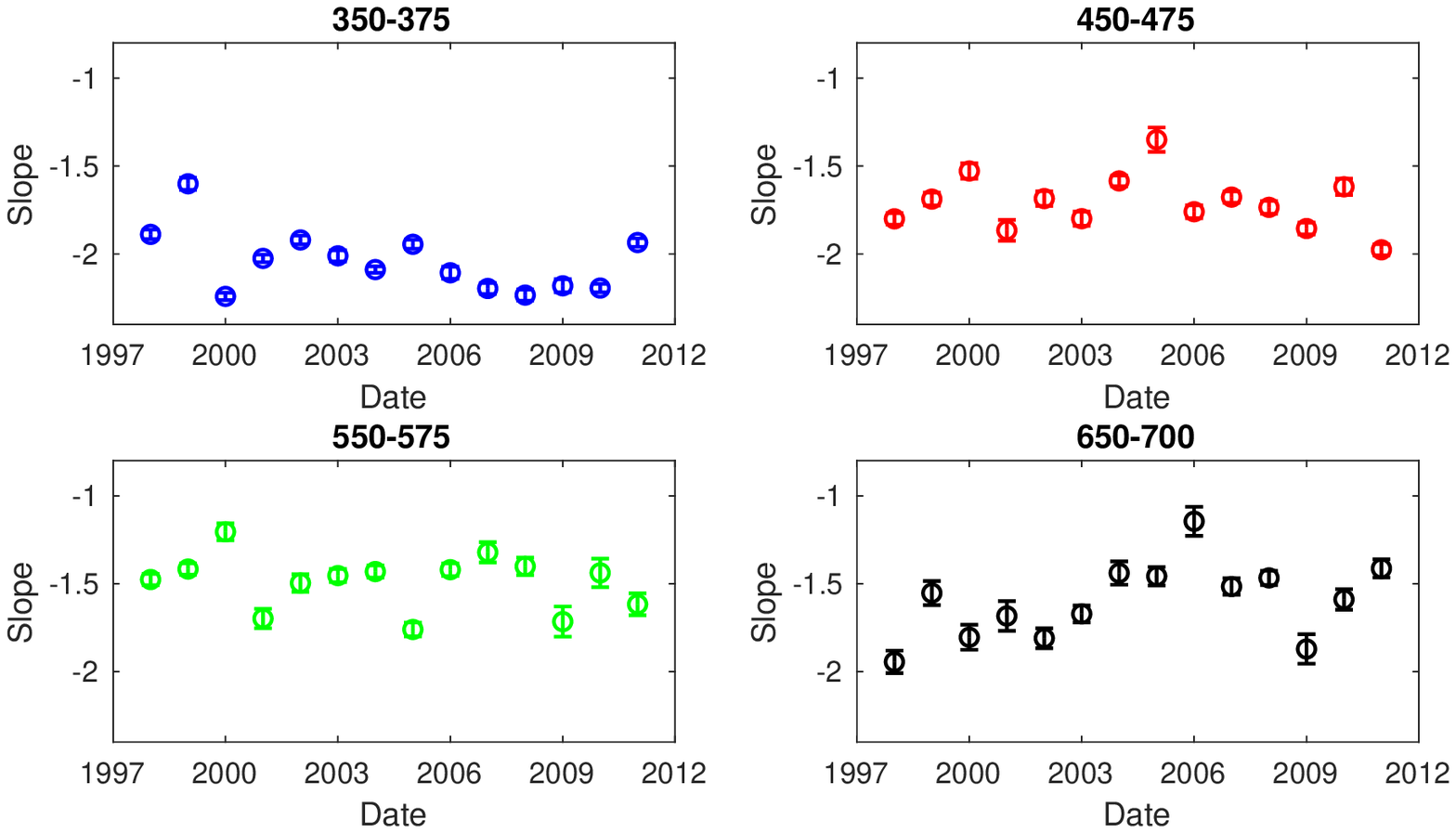}
\caption{{\bf Top two panels:} Annual averages of $\log T_{ion}$ vs the Z/A ratio, for 
four velocity classes in 2002 (first panel) and 2008 (second panel), along with a linear 
fit. {\bf Bottom two panels:} value of the $a_{Z/A}$ slope as a function of time along 
the solar cycle, for each velocity class shown in the top two panels (see text for details).
\label{z_a_fits}}
\end{figure}

Several heating mechanisms predict that ion temperatures depend on the charge-to-mass
ratio (Landi \& Cranmer 2009 and references therein). To test this dependence, and check 
whether such a dependence is affected by the solar cycle, we have fitted a relationship 
of the type 

\begin{equation}
Log~T_{ion} = a_{Z/A} {\left({\frac{Z}{A}}\right)}+b_{Z/A}
\end{equation}

\noindent
to the annual averages of $T_{ion}$ in each velocity bin. Examples of the fit for 
four select velocity bins in 2002 (solar maximum) and 2008 (solar minimum) are shown 
in Figure~\ref{z_a_fits} (top): the correlation coefficient is best at slower speeds, 
and decreases in the fast wind, although the fit is significant at all speeds. 
Figure~\ref{z_a_fits} (bottom) displays the $a_{Z/A}$ coefficient as a function of 
the solar cycle for each of the four velocity bins in the top panel: there is no 
indication that $a_{Z/A}$ correlates either with velocity, or with the solar cycle.

Figure~\ref{z_a_fits} (top) indicates that a few data points, marked in green, were 
discarded. These points were discarded when the difference between the average annual
$T_{ion}$ value and the fitted one was larger than 0.5~dex: after being removed, the
fit was re-done. These points correspond to three categories of ions: 

\begin{enumerate}

\item Ions like C$^{3+}$, Fe$^{19+}$ and similar, with less than 5 counts to calculate 
the annual average $T_{ion}$: these ions were considered too uncertain;
\item He$^{1+}$: this ion has a relatively large number of annual counts to calculate
an average, but Rivera \etal (2021) found that most of He$^{1+}$ may be formed in the
inner Heliosphere by charge exchange interactions between wind $\alpha$ particles and 
outgassed Helium and Hydrogen originating from circumsolar dust, so its properties may 
be influenced by different processes than all the other ions and for this reason was 
removed from the fit; 
\item He$^{2+}$: the annual average $T_{ion}$ for this ion is {\em consistenly} lower 
than the fitted value by amounts of the order of 0.5~dex (so that sometimes it was not 
removed from the plot). We will further discuss this He$^{2+}$ feature in 
Section~\ref{collisional_age}.

\end{enumerate}

\subsubsection{Ion temperature vs mass}

We also checked whether the ion-to-proton temperature ratio is mass-proportional,
and whether such a dependence changes with the phase of the solar cycle. We found 
that no significant dependence was present. The ratio $T_{ion}/T_p/M_{ion}$ ranges 
between 1 and 2 for all the ions we considered with no specific trend. Within 
each element, different ionization stages had a different value in this range, 
consistently with the clear dependence of the $T_{ion}/T_p$ ratio on the Z/A ratio, 
but no trend was identified neither with ionization stage, nor with time along the 
solar cycle.

\subsection{Differential velocity}

\begin{figure}
\includegraphics[height=9.0cm,width=18.0cm]{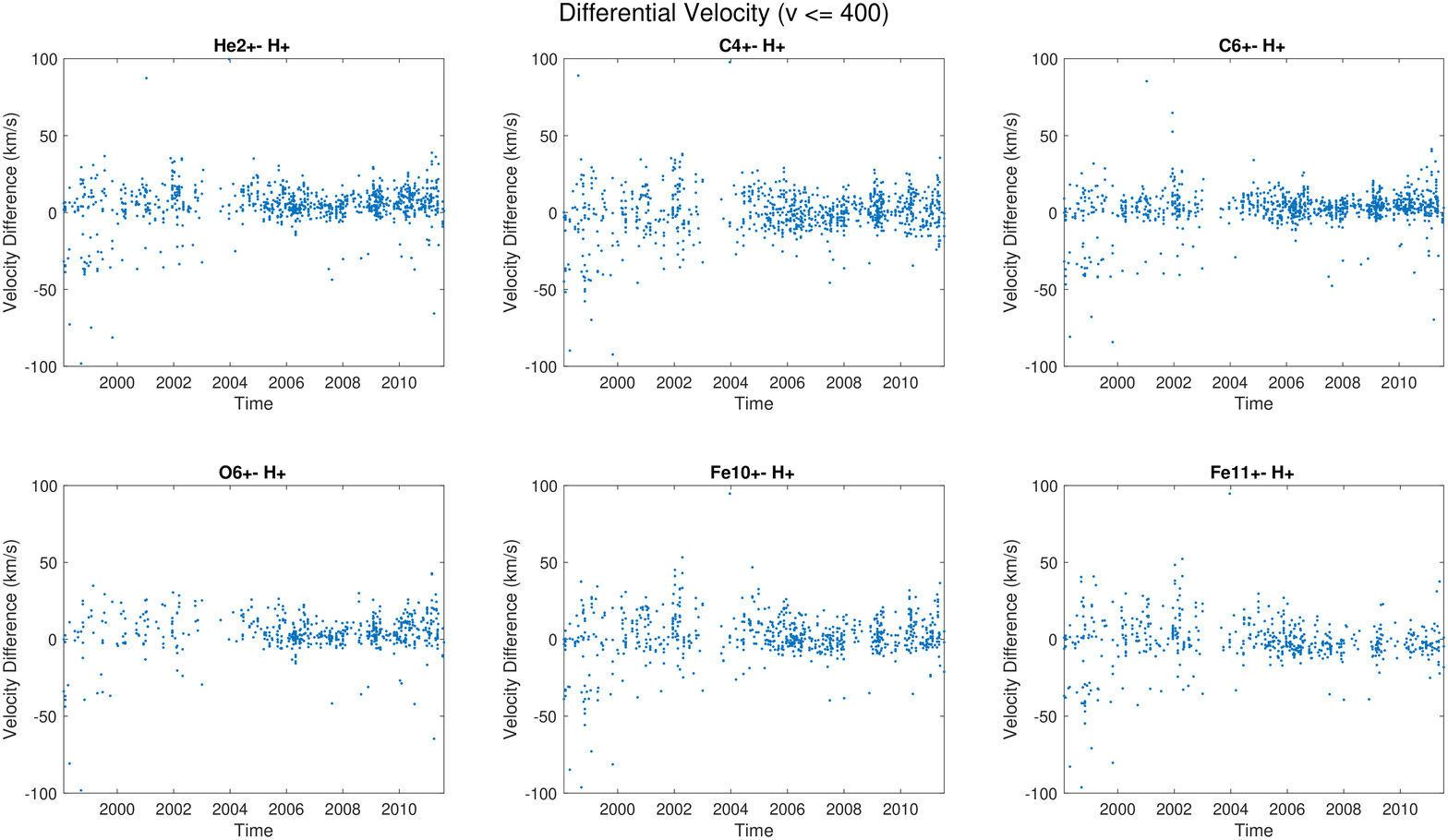}
\includegraphics[height=9.0cm,width=18.0cm]{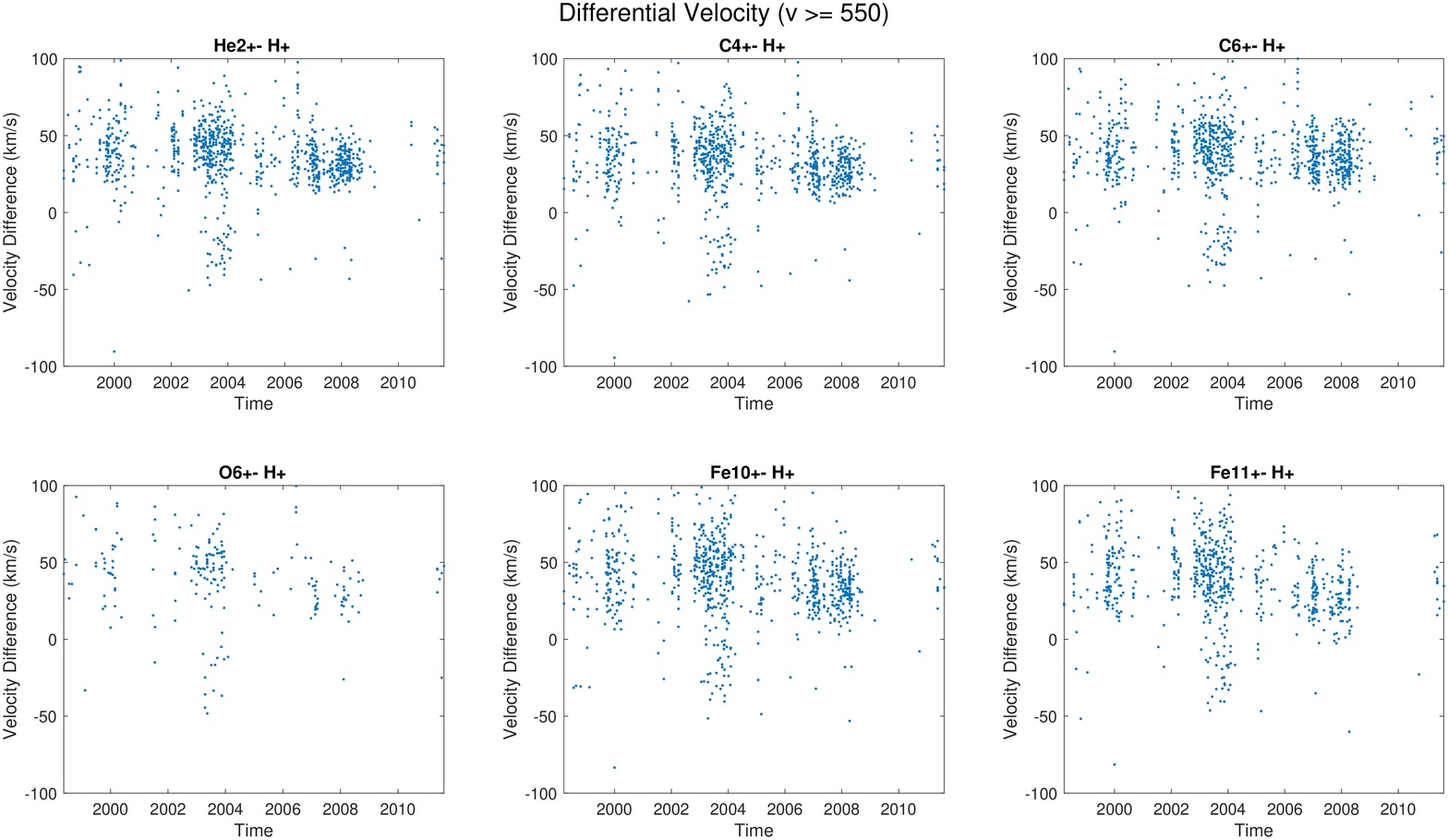}
\caption{Differential velocities for a selection of ions along the solar cycle for
velocities slower than 400~\kms (top two rows) and faster than 550~\kms (bottom two
rows).
\label{diff_velocity}}
\end{figure}

\begin{figure}
\includegraphics[height=9.0cm,width=18.0cm]{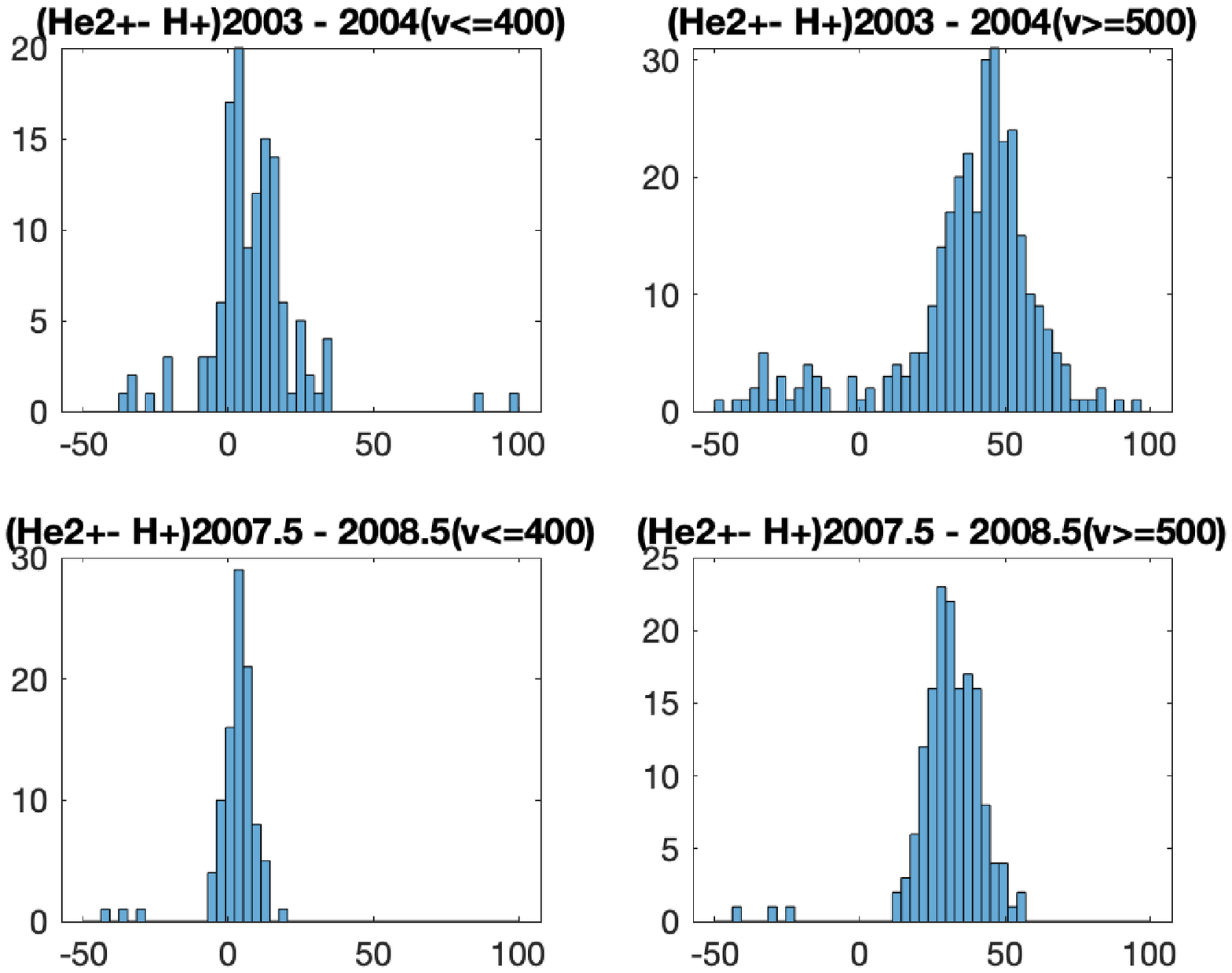}
\includegraphics[height=9.0cm,width=18.0cm]{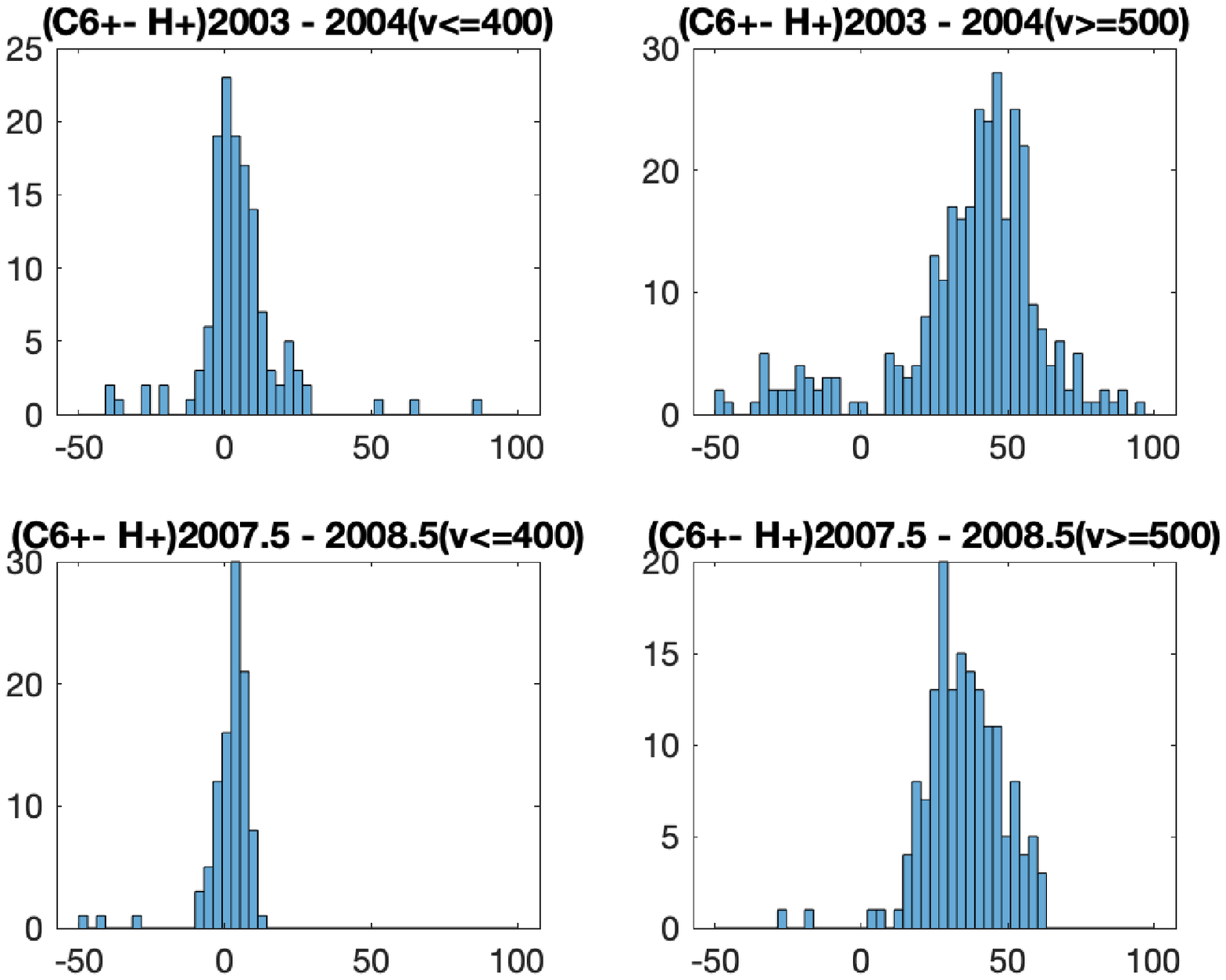}
\caption{Differential velocities for a selection of ions along the solar cycle for
velocities slower than 400~\kms (top two rows) and faster than 550~\kms (bottom two
rows). Results are shown at solar maximum (first and third row) and minimum (second
and fourth row).
\label{diff_velocity_hist}}
\end{figure}

Ion-proton differential velocity was measured using only plasma parcels where the 
radial component of the magnetic field was 90\% or more of the total magnetic field 
magnitude, and that the magnetic field variability was limited within each 2-hour
timebin. These requirements greatly decreased the number of data, so that resolution 
in velocity needed to be sacrificed in order to preserve a sufficient number of 
counts to let any trend be discernible. We divided the available data in two 
velocity classes, slow ($v<400$~\kms), and fast ($v>550$~\kms) in an effort to isolate 
fast and slow wind streams; even so, the number of available data points is very 
small. Results are shown in Figure~\ref{diff_velocity}, and example of histograms 
for He$^{2+}$ and C$^{6+}$ are shown in Figure~\ref{diff_velocity_hist}.

Two main conclusions can be drawn from the data. At velocities slower than 400~\kms, 
differential velocity is smaller than 10-15~\kms~for all ions, and does not show any 
dependence on the solar cycle, remaining constant in this interval. At velocities 
larger than 550~\kms, speed differences range between 20 and 70~\kms, with only a 
moderate decrease during solar minimum. Furthermore, the lack of data did not allow 
us to determine whether there is any dependence of the differential speed on the 
ions' charge, mass, or charge/mass ratio. The presence of a small population of 
measurements with negative speeds (e.g. ions are slower than protons) around 2004 
for the fast wind class is a puzzle: we have investigated whether they could be due 
to magnetic switchbacks (Yamauchi \etal 2004) but an inspection of MAG data in the 
proximity of these points did not reveal a systematic presence of such magnetic 
structures.

\subsection{Charge state composition}

\begin{figure}
\includegraphics[height=9.0cm,width=18.0cm]{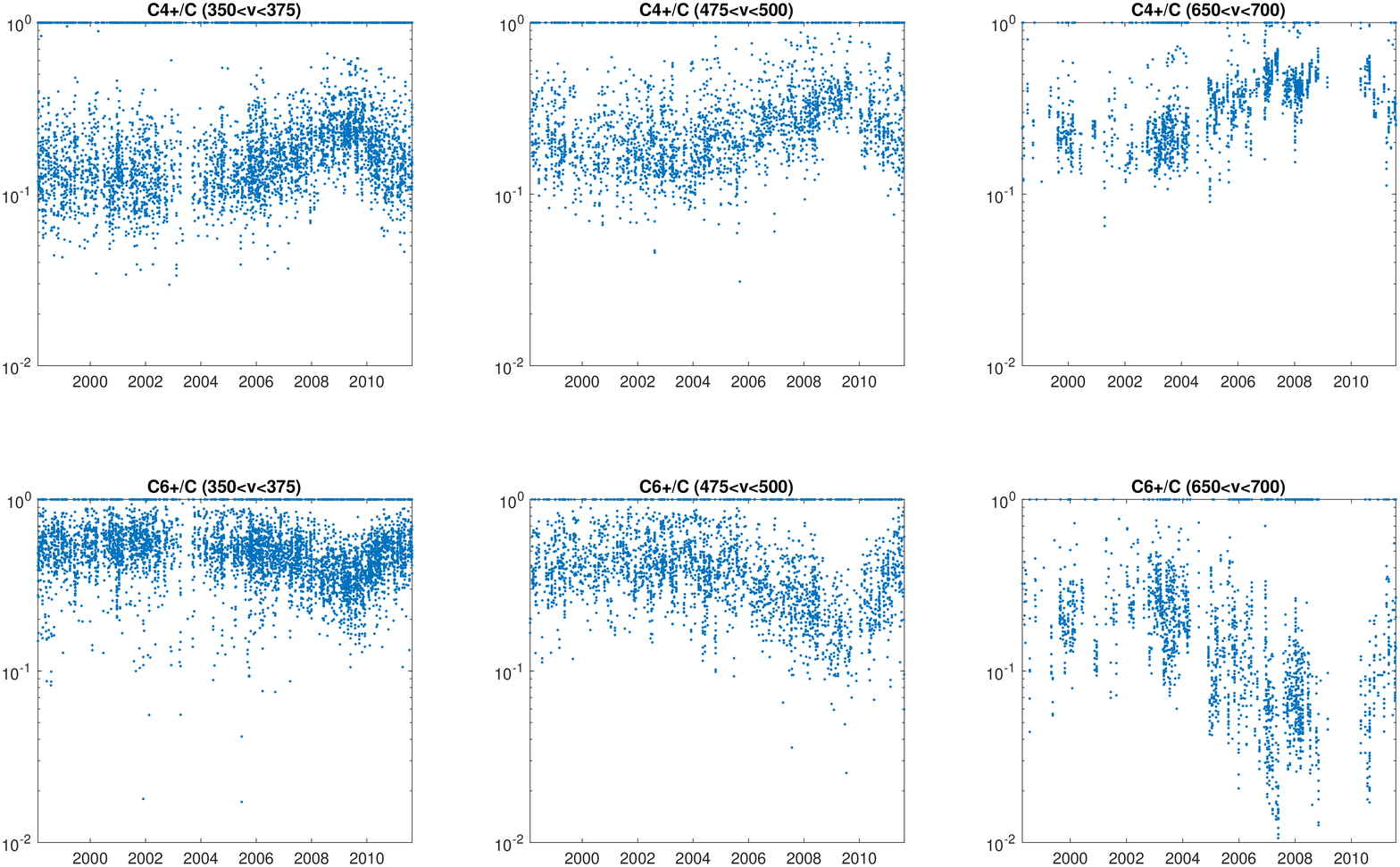}
\caption{Relative abundances of a selection of C ions along the solar cycle, for three
velocity classes ranging from slow, intermediate, and fast wind.
\label{charge_states_1}}
\vspace{0.5cm}
\includegraphics[height=9.0cm,width=18.0cm]{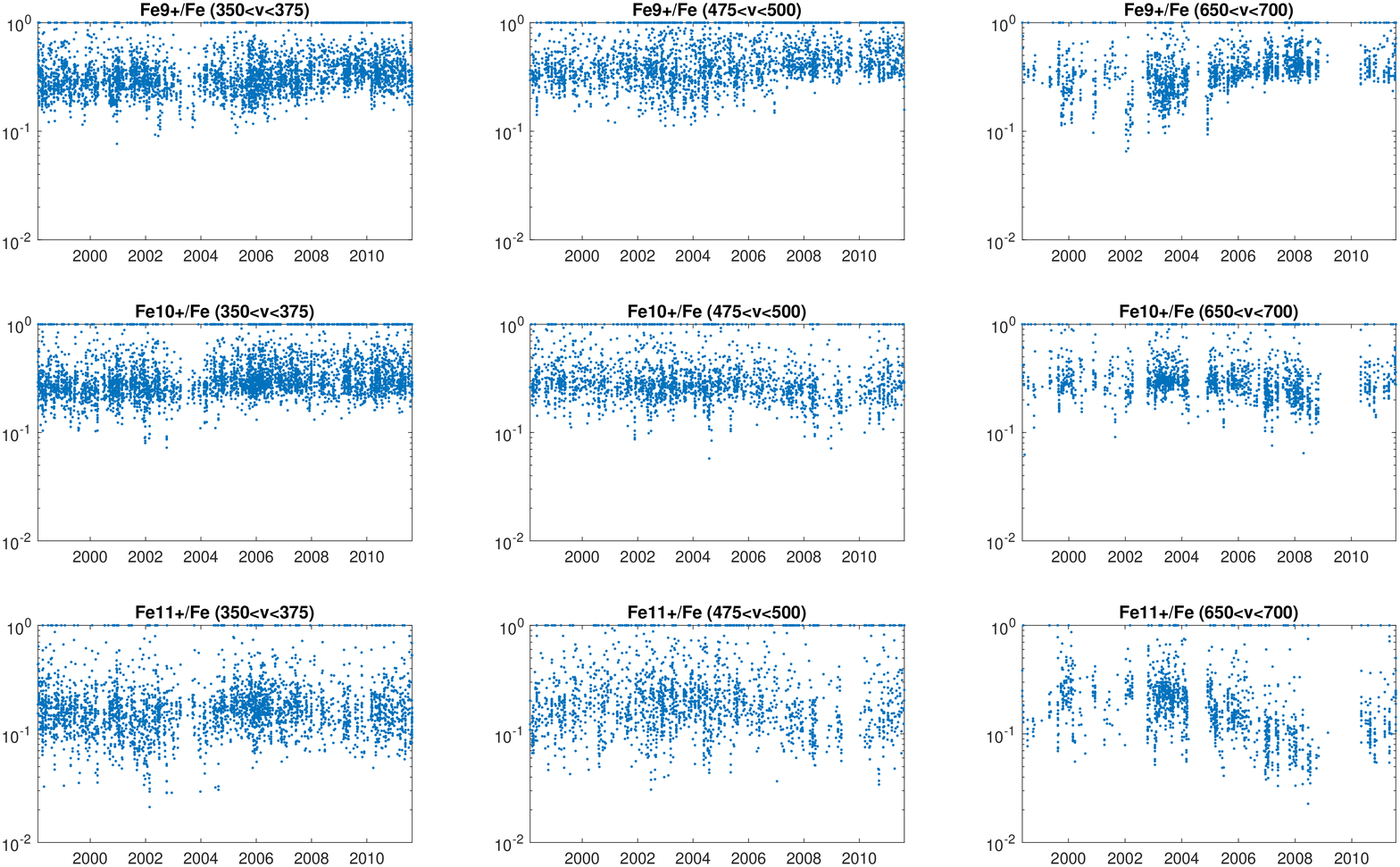}
\caption{Relative abundances of a selection of Fe ions along the solar cycle, for three
velocity classes ranging from slow, intermediate, and fast wind.
\label{charge_states_2}}
\end{figure}

\begin{figure}
\includegraphics[height=10.0cm,width=18.0cm]{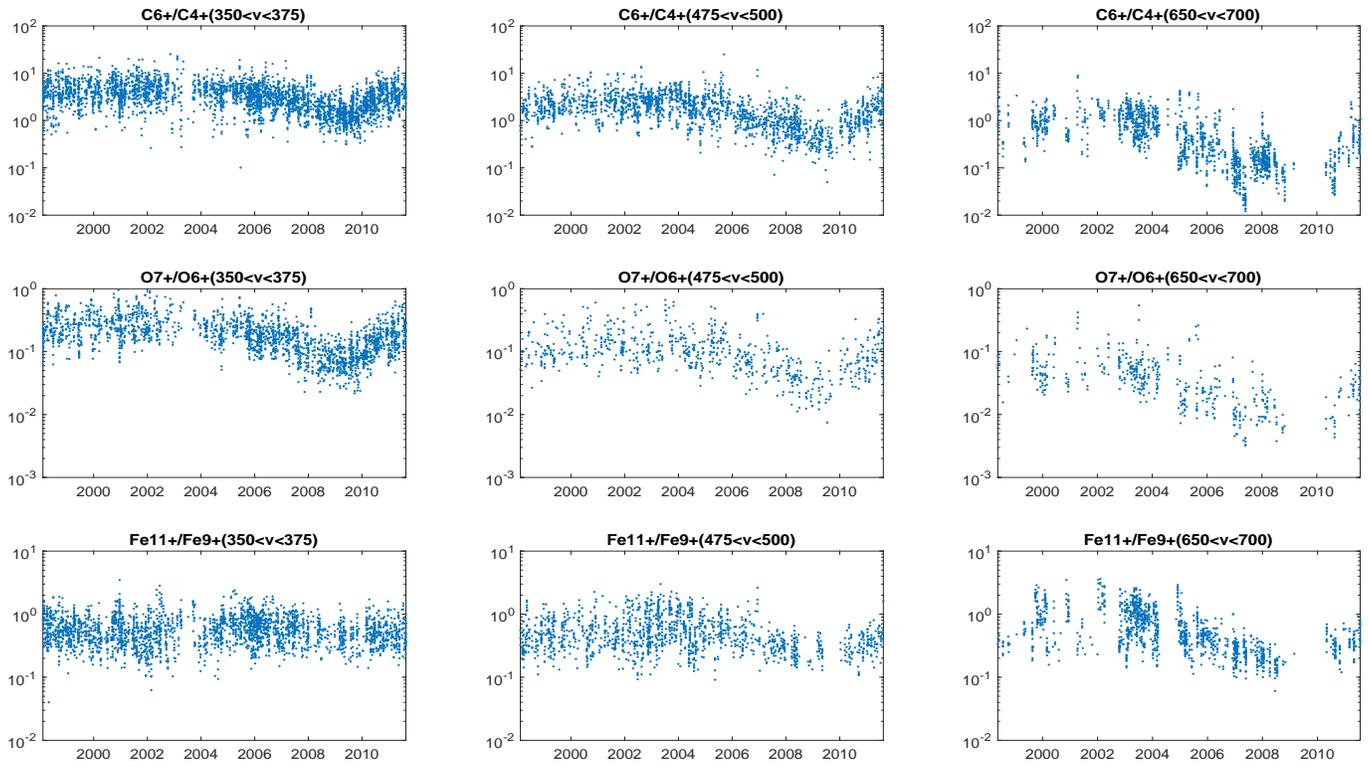}
\caption{Select C and O charge state ratios along the solar cycle, for three
velocity classes ranging from slow, intermediate, and fast wind.
\label{charge_state_ratios}}
\end{figure}

Charge state composition strongly depends on the solar cycle, with all elements and 
ions showing higher ionization during maximum than during solar minimum with the only 
exception of Neon ions. A few examples are shown in Figures~\ref{charge_states_1} 
and \ref{charge_states_2}. As a consequence, charge state ratios show an even more 
marked dependence on the solar cycle, as shown in Figure~\ref{charge_state_ratios}. 
This behavior confirms the cycle-related variability of a few key charge state ratios 
noted by Lepri \etal (2013). It is important to note that such a behavior is also 
clearly shown by individual Fe ions and Fe charge state ratios, even though the average 
charge state of Fe 
does not strongly depend on the solar cycle: even though ionization changes from 
solar minimum to maximum are not as large as for Carbon and Oxygen ions, they are 
significant and indicate that the ionization status of Fe is also significantly 
changing with time. This last result shows that using the average charge states 
to characterize the ionization status of Fe dilutes the effects of the solar 
cycle on this element.

\subsection{Low collisional age plasma}
\label{collisional_age}

Proton-ion Coulomb collisions tend to equalize ion temperatures and differential 
velocity signatures in the solar wind, while they have no effect on charge state 
composition. Thus, restricting the analysis only to wind stream with collisional 
age $A_C< 0.3$ allows us to investigate more pristine solar wind streams, whose 
properties are closer to those much closer to the Sun. 

However, enforcing the $A_C<0.3$ limit to the data means that very few measurements 
were left in any velocity bin slower than 350~\kms, and that the number of surviving 
data is also reduced below 500~\kms. At larger speeds, all the trends we identified 
in the complete data set are confirmed, because removing the high collisional age 
streams did not affect the data set significantly since the plasma density is 
sufficiently low to always satisfy the $A_C< 0.3$ criterion with the only exception
of the densest streams.

At lower speed instead, the surviving data were too few to allow any investigation
on differential velocity. On the contrary, sufficient data were left to repeat our 
study of the ion temperatures.

\subsubsection{Ion temperature values}

\begin{figure}
\vspace{3mm}
\includegraphics[height=7.5cm,width=16.0cm]{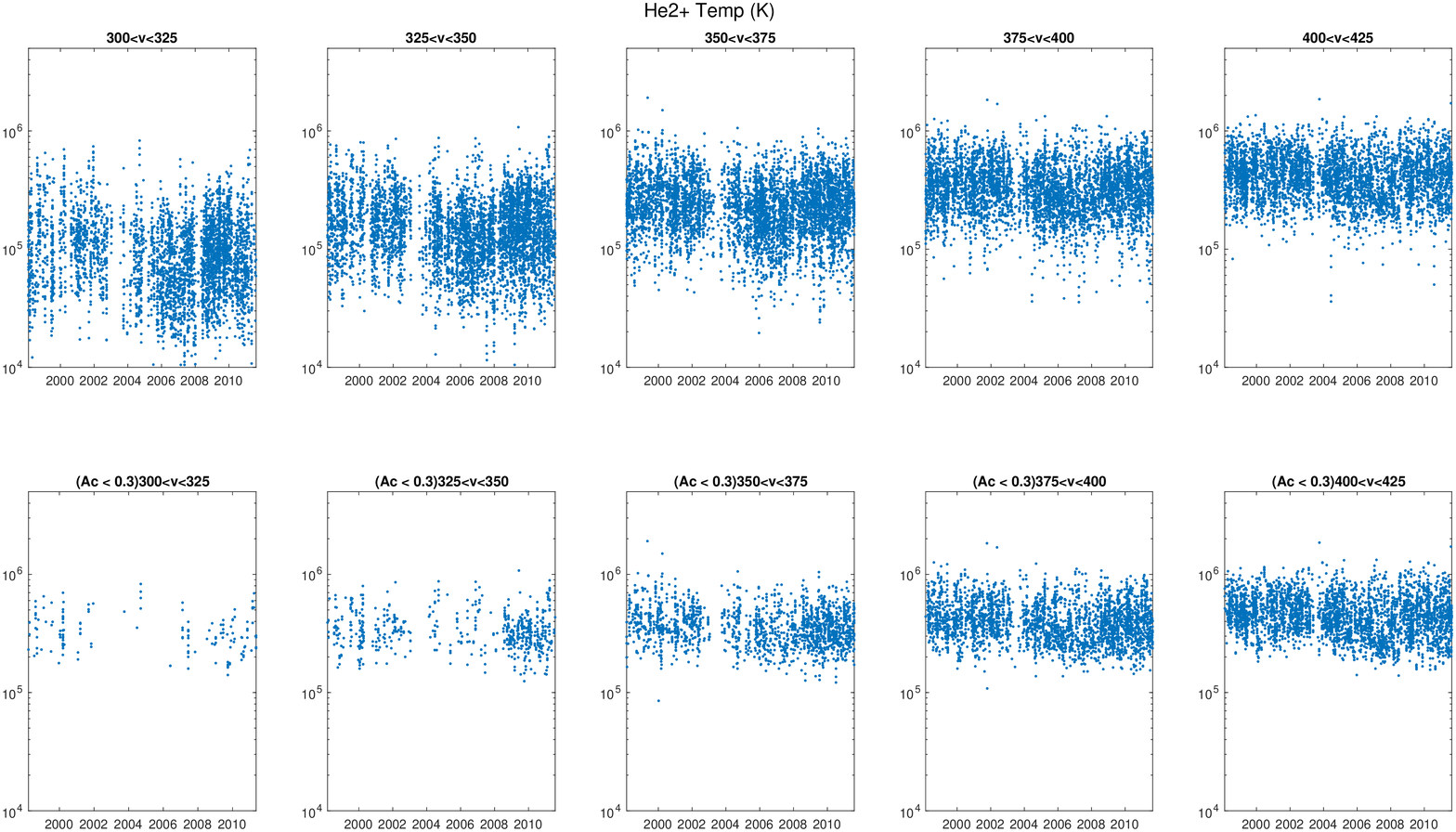}
\vspace{3mm}
\includegraphics[height=7.5cm,width=16.0cm]{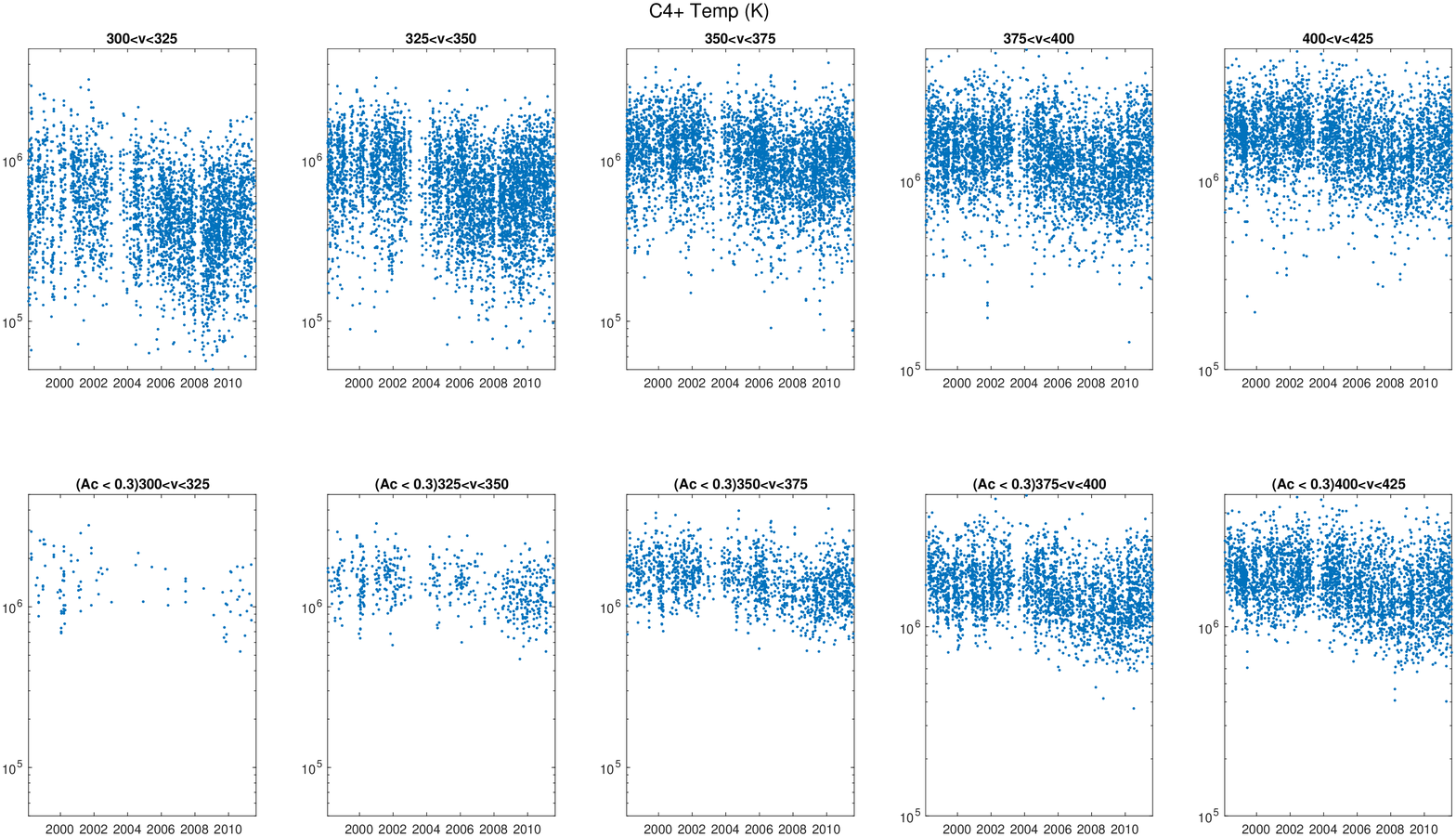}
\vspace{3mm}
\includegraphics[height=7.5cm,width=16.0cm]{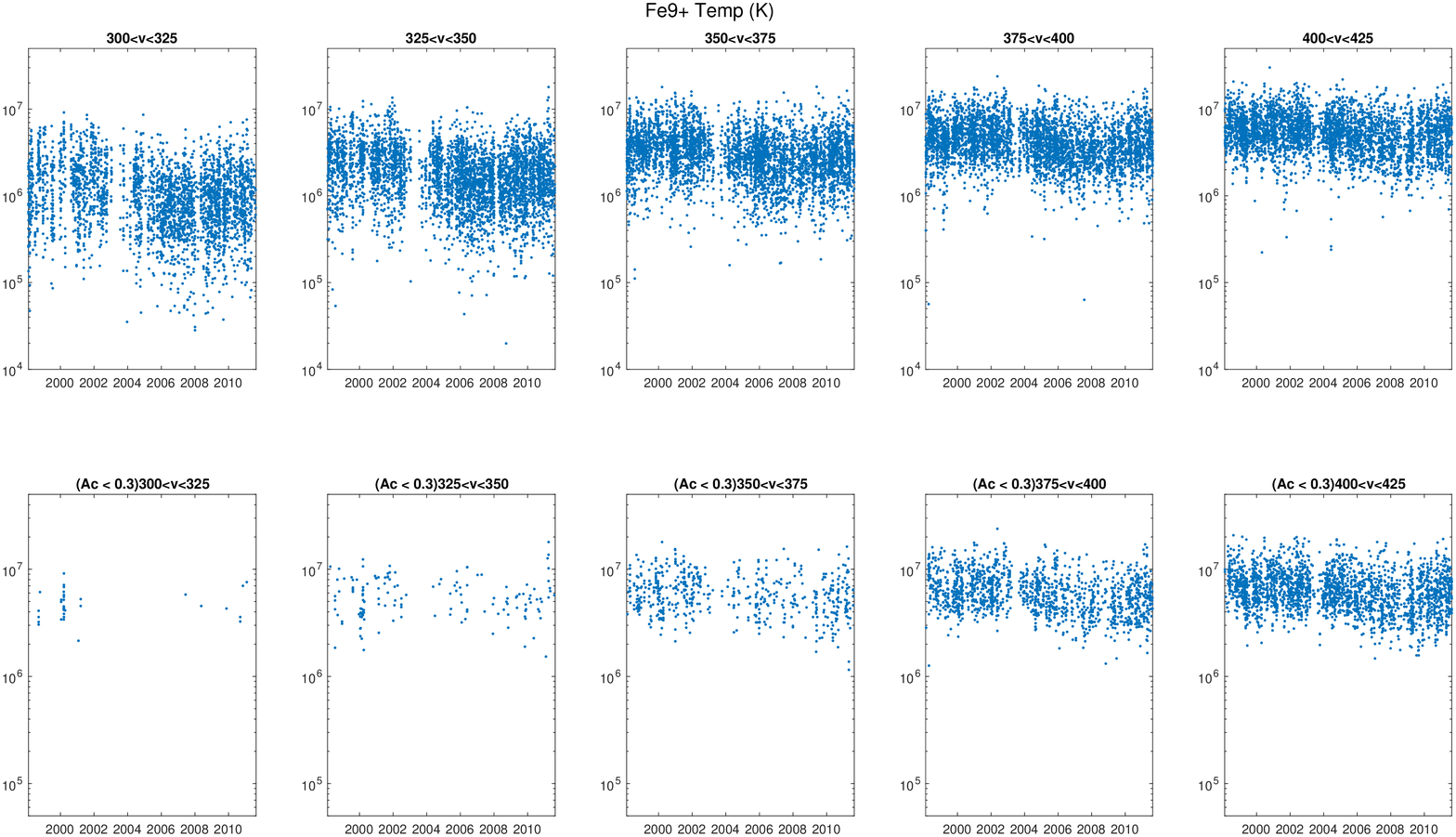}
\caption{Comparison of ion temperatures at low speeds in collisionally
young wind and the complete data set. First two rows: He${2+}$, middle
two rows: C${4+}$, bottom tworows: Fe$^{9+}$. For each ion, the general
data set is shown on top, and the collisionally young data set is shown
on bottom.  
\label{tion_low_ac}}
\end{figure}

Figure~\ref{tion_low_ac} shows the effect of removing all collisionally old
data from the sample, for a few representative ions: He$^{2+}$, C$^{4+}$ and Fe$^{9+}$.
Only speeds bins slower than 425~\kms are shown, to better illustrate the effects of 
Coulomb collisions; for each of the ions, the top row shows all the data in the
bin, and the bottom row only the collisionally young plasma parcels.

Figure~\ref{tion_low_ac} shows three main results:

\begin{enumerate}

\item As expected, the slower the speed bin, the higher the density, so that increasingly
less streams are collisionally young; 

\item While the full dataset shows an overall large decrease of the ion temperatures
as the speed of each bin decreases, the collisionally young data show very limited to
no decrease in ion temperature;

\item In collisionally young velocity bins with enough surviving data, the dependence 
of the ion temperature on the solar cycle is preserved, indicating that it is likely
not due to Coulomb collisions, but to a real process taking place in the solar wind.

\end{enumerate}

\subsubsection{Ion temperature vs charge-to-mass ratio and wind speed}

\begin{figure}
\includegraphics[height=8.0cm,width=16.0cm]{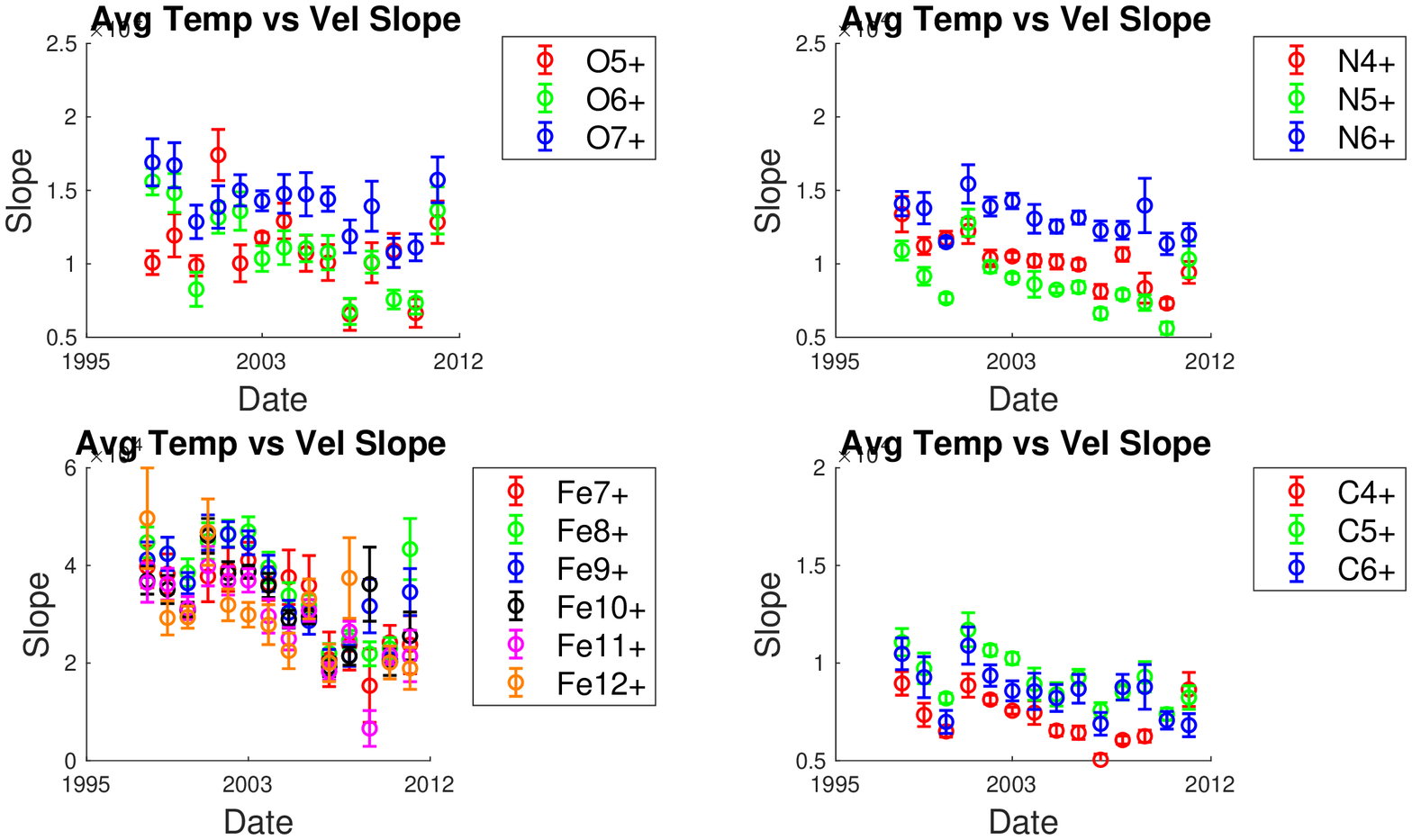}
\includegraphics[height=6.0cm,width=8.0cm]{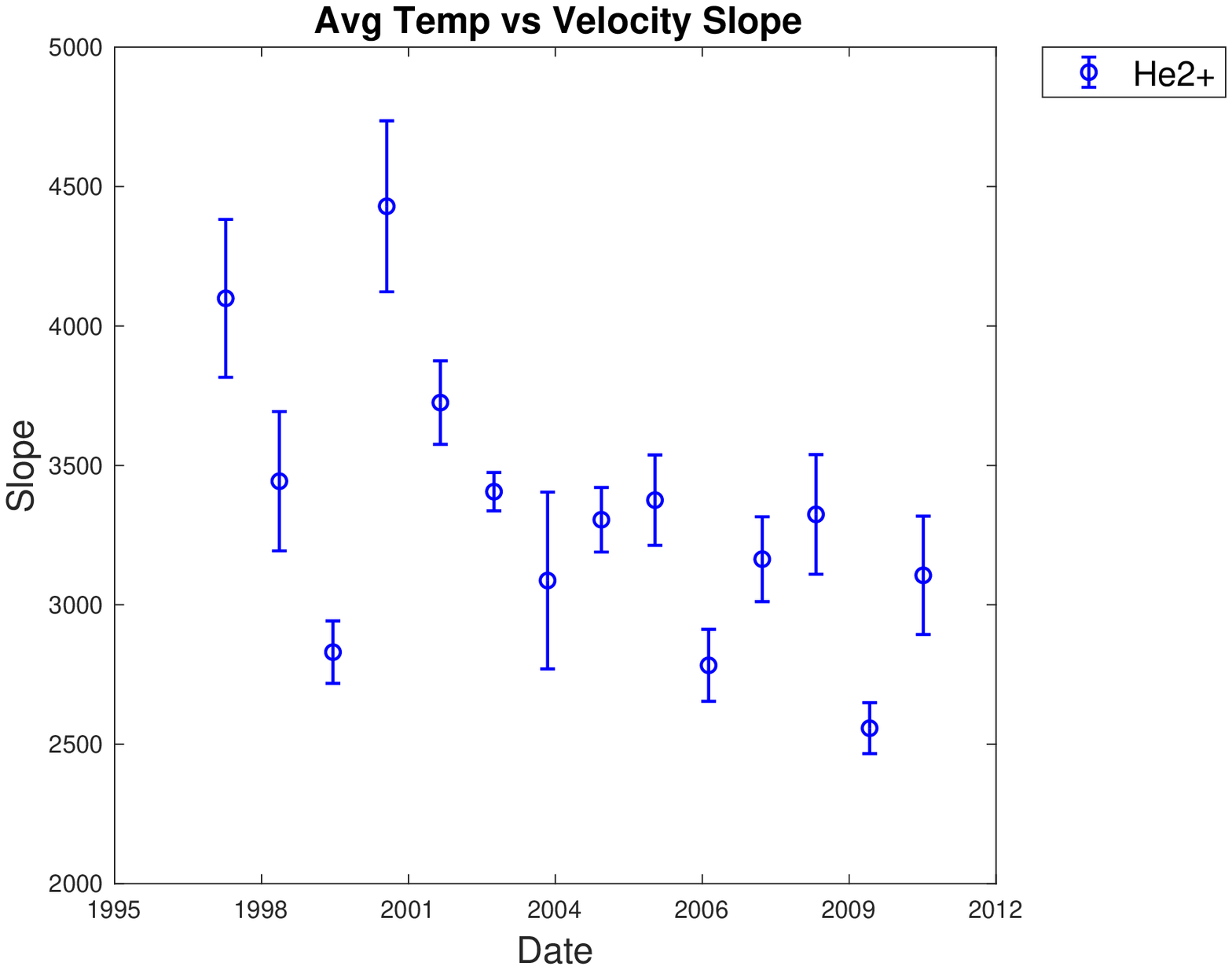}
\includegraphics[height=6.0cm,width=8.0cm]{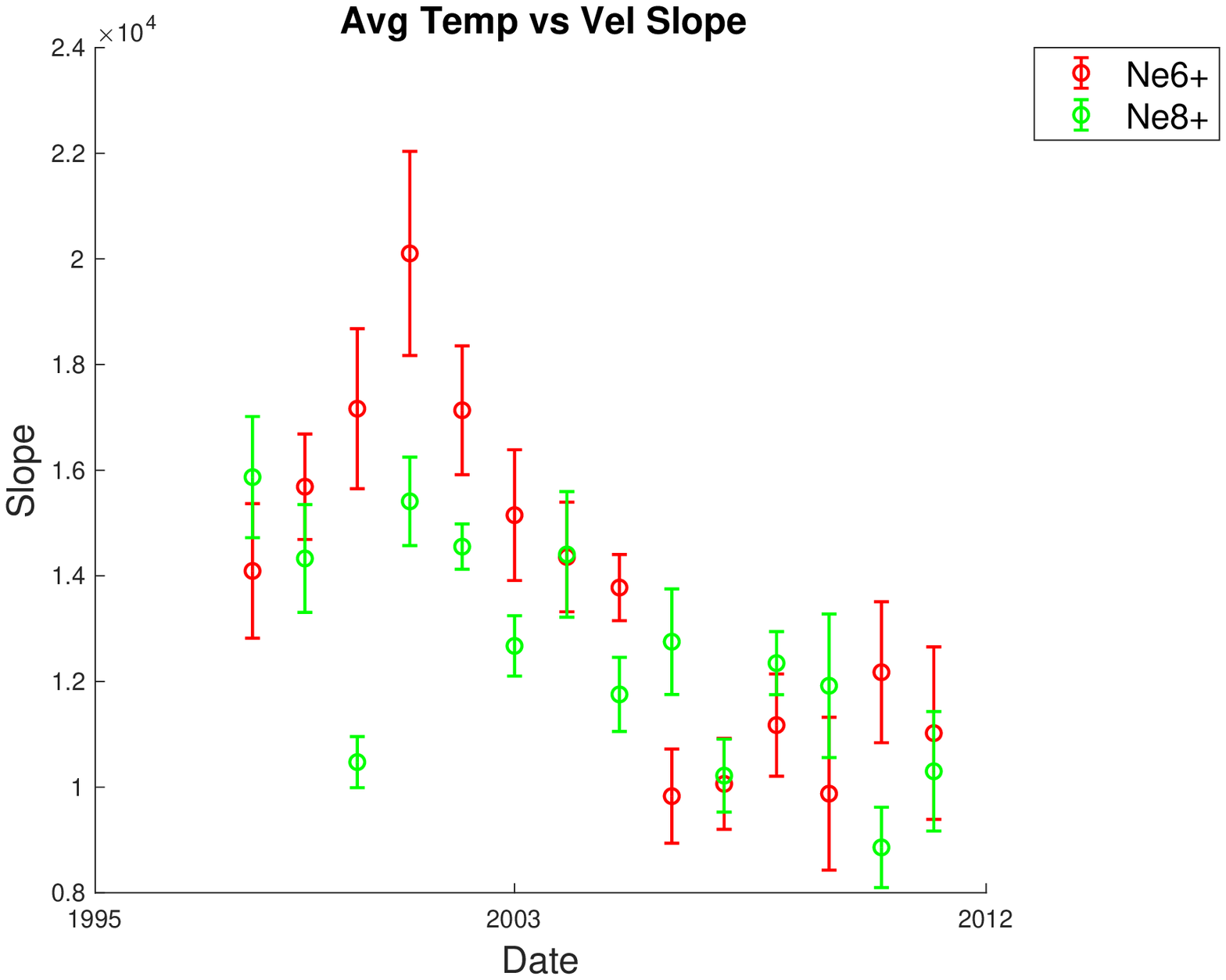}
\caption{Same as Figure~\ref{tp_vs_vel}: values of $a_{lin}$ for the
collisionally young wind. 
\label{tion_za_low_ac}}
\end{figure}

The larger ion temperature values of low-speed, collisionally young plasma question 
the results obtained fitting a linear relationship between ion temperature and speed
to the full dataset. Also, the relationship between ion temperature and the Z/A ratio 
may change. We have repeated those fits using only the collisionally young plasma. 
Figure~\ref{tion_za_low_ac} shows that the slope $a_{lin}$ of the linear fit between 
$T_{ion}$ and wind speed is:

\begin{enumerate}

\item shallower for all ions, consistently with slower wind having higher $T_{ion}$ values;
\item maintaining the same trend as a function of the solar cycle as the complete dataset,
being lower at minimum and higher at maximum;

\end{enumerate}

\noindent
Thus, although the values of $a_{lin}$ change, the overall solar cycle dependence 
determined with the full dataset is preserved also for the collisionally young one.

\begin{figure}
\includegraphics[height=8.0cm,width=16.0cm]{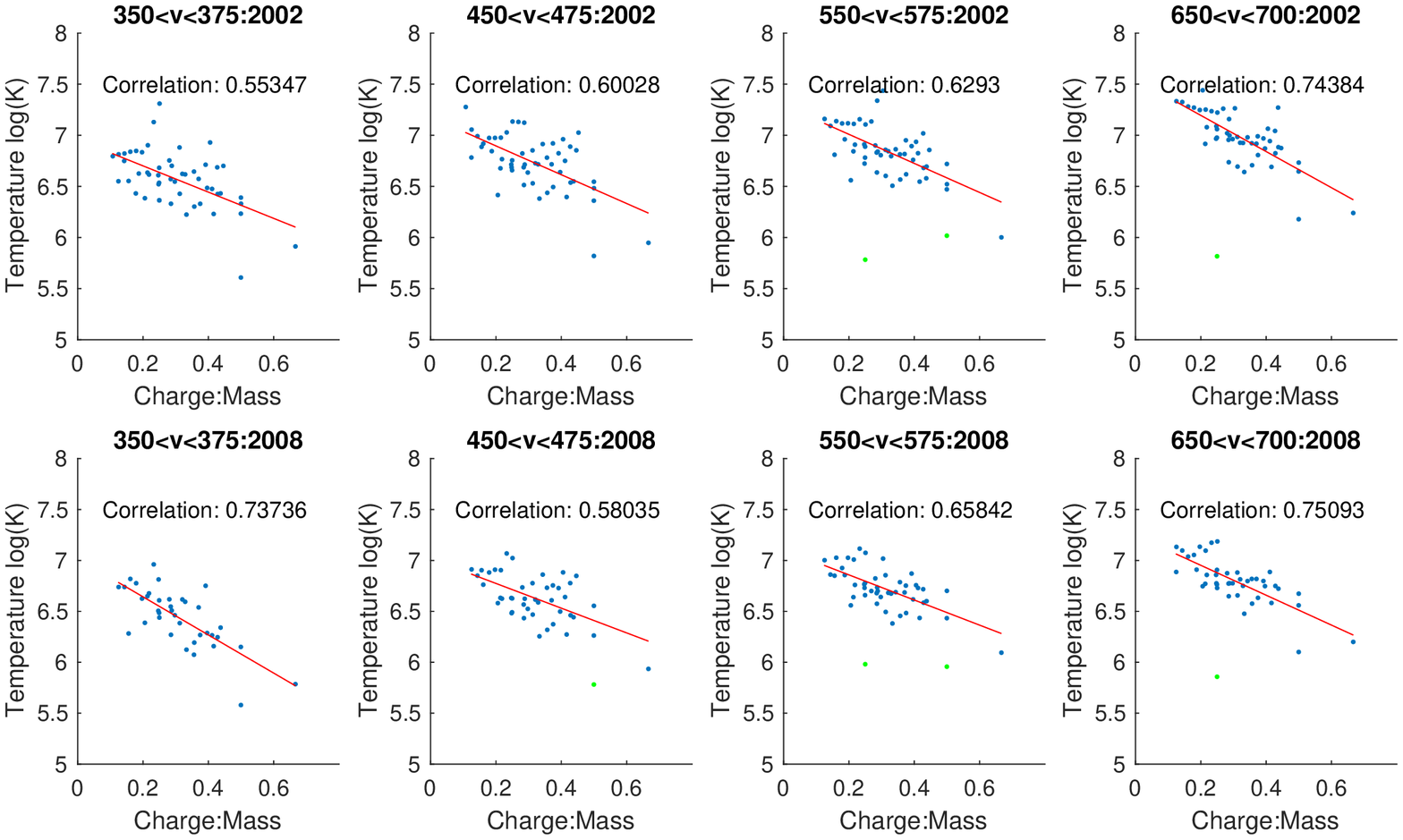}
\includegraphics[height=8.0cm,width=16.0cm]{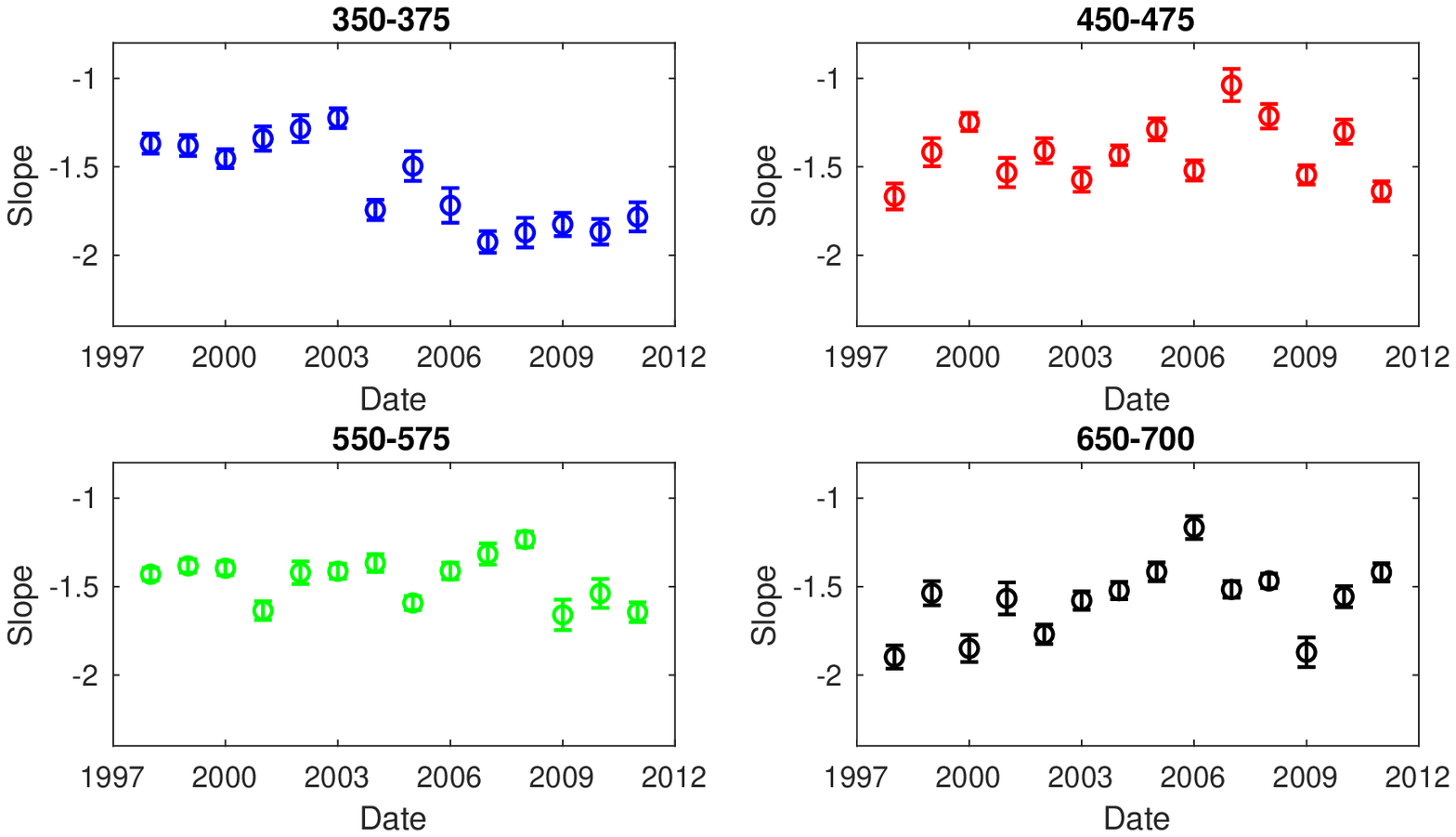}
\caption{Same as Figure~\ref{z_a_fits}: values of $a_{Z/A}$ for the
collisionally young wind.
\label{tion_za_low_ac_2}}
\end{figure}

As fas as the slope $a_{Z/A}$ of the linear fit between $\log T_{ion}$ and Z/A,
results are shown in Figure~\ref{tion_za_low_ac_2}. We find that $a_{Z/A}$ is:

\begin{enumerate}

\item unchanged at larger speeds (as expected); 
\item less steep at lower speed, and with the same scatter as the other speed bins;
\item having approximately the same value in all speed bins;
\item uncorrelated to the solar cycle.

\end{enumerate}

\noindent
These results indicate that the process generating a dependence of the ion temperature 
on the Z/A ratio is the same for all wind speeds, and does not change along the solar 
cycle.

\section{Discussion}
\label{discussion}


\subsection{Plasma density and wind ionization properties}


The long-term decrease of solar wind proton density, visible at all speeds except 
the slowest bins, does not seem to correlate with the solar cycle. Such a decrease 
has an unknown origin; Zerbo \& Richardson (2015) showed that 27-day proton density 
averages of the ecliptic solar wind started this decrease around 1990, and on longer 
timescales (dating back to the late 1960s) proton density changed slowly with timescales 
longer than 40 years. Also, we find that the proton density decrease does not affect the 
well known density-vs-velocity relationship: a double-sloped linear fit of yearly 
averages of the proton density with speed shows little variability, indicating that
the measured density decreased affects all winds in the same way.


This slow, cycle-unrelated density variation should result in a gradual decrease of 
both ionization and recombination rates due to free electron-ion collisions, as the 
overall density of free electrons follows the proton density. In turn, decreased 
ionization and recombination should lead to freeze-in heights increasingly closer 
to the Sun and to lower wind ionization. Furthermore, as for the proton density, 
these changes should not show a clear correlation with the 11-year solar cycle. On the 
contrary, the present results confirm the findings of Lepri \etal (2013), who noted 
that the plasma charge state distribution closely tracks the solar cycle, being lower 
at minimum, and increasing together with the sunspot number. Most importantly, both 
the present results and the Lepri \etal (2013) results point towards an increase in 
wind ionization after 2010, contrarily to the expectations from the ongoing decrease 
of the electron density. This means that electron density is not the only cause of 
wind ionization evolution.

Landi \& Lepri (2015) attempted to explain the charge state variations with
photoionization from solar EUV and X-ray radiation outside flares, as EUV and X-ray
fluxes are mostly emitted by active regions and depend critically on the solar cycle. 
Landi \& Lepri (2015) found that photoionization indeed induces some variability, 
but it was insufficient to explain the measured variation in charge state ratios. 
Landi (2022, in preparation) added flare photoionization to the picture, but found 
that flare-enhanced high-energy photoionizing flux could not account for the missing, 
cycle dependent ionization. Cycle-related changes of plasma density and temperature 
in quiescent streamers, which might be responsible for ionization variations at least 
in the slow solar wind, were ruled out by Landi \& Testa (2014) using remote sensing 
observations. Systematic measurements of density, temperatures and emission measures 
in equatorial coronal holes -- the source of a large fraction of the fast speed wind 
observed by ACE -- were carried with SDO/AIA from 2010 to 2019 encompassing most of 
cycle~24, and they also showed no significant variation along the solar cycle 
(Heinemann \etal 2021). Thus, the reason behind the solar cycle variation of wind 
ionization, which affects all elements, is still unclear. 

Wind ionization is determined by plasma electron temperature, electron density, and 
speed. The first two quantities determine the local collisional ionization and
recombination rates at any location along the wind trajectory, while speed determines
the time spent by a wind plasma parcel at each location. Since electron density and
temperature seem not to be the cause of the observed ionization variations, the only 
remaining possibility seems to be wind acceleration close to the Sun. In order to 
cause lower ionization at solar minimum, plasma speed needs to be faster close to 
the Sun, so that the solar wind spends less time in the densest and hottest part 
of the corona. This means that solar wind acceleration needs to be more efficient 
close to the Sun during solar minimum and less efficient at solar maximum. However, 
no independent confirmation of this scenario is currently available with remote 
sensing measurements.

Wind ionization models rely on ionization and recombination rates calculated using
a Maxwellian distribution of electron velocity. Tails of high-energy electrons can
significantly increase ionization rates, leading to a higher wind charge states 
(Cranmer 2014): in case their presence and size depended on the solar cycle, they 
could account for the observed cycle variations of wind ionization. Unfortunately, 
these high-energy electron populations have not yet been unambiguously identified 
by remote sensing observations.

\subsection{Plasma fractionation processes: absolute and relative elemental abundances}

Elemental abundances open a window on the evolution of the FIP effect along the solar 
cycle, which is largely unavailable from remote sensing observations due to the lack 
of a comprehensive and continuous monitoring of the solar corona with spectrally 
resolved data. 

\subsubsection{Absolute abundances}

The variability of absolute abundances indicates a change in the capability of wind 
acceleration mechanisms to extract heavier ions over protons from the wind's source 
region. The present results indicate that the composition of the slow solar wind is 
sensitive to the solar cycle, with the amount of heavy ions present in this wind being 
lower during solar minumum relative to solar maximum. This result affects all elements, 
including Oxygen, usually assumed to track Hydrogen's response to ponderomotive forces,
and confirms the results obtained by Lepri \etal (2013). At faster speeds such a solar 
cycle dependence is much reduced. 

These results may be due to two different scenarios. First, the decreasing amount 
of change in composition as the wind speed approaches 500~\kms~is consistent with a 
"two-wind scenario" where a coronal hole-associated wind is present at all speeds 
whose metallicity is weakly affected by the solar cycle, and a reconnection-driven 
slow wind is present at speeds lower than 500~km/s, which comes from initially closed 
magnetic structures and whose metallicity is lower at solar minimum. The decreasing 
amount of solar cycle dependence as speed approaches 500~km/s may be simply due to 
a decreasing fraction of wind originating from closed loop structures.

Second, if the observed FIP effect variations are due to the slow wind component
coming from formerly closed loops, the present results may also suggest a marked 
change in the FIP fractionation occurring in closed magnetic structures as the
solar cycle unfolds, which does not occur in open field lines.

The open question is to what extent can the FIP variation in the slow wind be 
ascribed only to a varying mixture in the wind source region, or to variability 
in the closed loops before their opening by reconnection, or both.

\subsubsection{Relative elemental abundances}

Similar results are found when relative abundances are considered. In fact, the 
FIP bias of the fast wind is largely constant and ranges between 1 and 1.5,
consistently with model predictions (Laming 2015). Slow wind FIP bias, on the
contrary, decreases during solar minimum, at which time the relative abundances
in the slow wind resemble those in the fast wind. 

As for the absolute abundances, this result is also consistent with the two-wind 
scenario, and it further suggests that coronal-hole associated wind dominates at 
all speeds during minimum, and that the fractionation process in its source region 
(coronal holes) is relatively stable during the solar cycle. At solar maximum slow
wind coming from closed loop structures, whose FIP bias can even exceed 10, becomes 
prevalent at slow speeds. Thus, the variability in slow wind FIP fractionation may 
again be due either to an evolving mixture of source regions, or to a changing 
effectiveness of FIP fractionation in closed loops, or both.

Looking at individual abundance ratios, the relative C/O abundance is remarkably 
stable along the solar cycle at all speeds, indicating that the process that 
increases its value from the photospheric value of 0.45-0.55 to the measured 
$\approx$0.7 does not change in time. According to the ponderomotive force model 
(Laming 2015 and references therein) the $\approx$1.3-1.6 enhancement in the 
C/O ratio from open magnetic flux tubes in coronal holes can be due to slow-mode 
wave amplitudes of around 6.0~\kms. The stability of this ratio in time indicates 
that this process is not affected by the solar cycle, and its uniformity over all 
velocity ranges suggests that it is ubiquitous in all source regions of the solar 
wind, even if the measured C/O ratio enhancement is slightly larger than predicted
for closed loop structures. The stability of this ratio in time argues for a
scenario where the varying abundances in the slow wind are only due to a changing 
mixture of source regions at different phases of the solar cycle.

The relative Fe/C abundance, which can be taken as a proxy for the low-FIP/high-FIP 
relative abundance, shows two interesting results. First, its value ranges over a 
much larger span in the slow wind than in the faster wind, as shown in the upper 
row of Figure~\ref{abundance_ratio_histogram}. This is again consistent with the 
"two-wind" scenario, and indicates that the relative abundance of wind released 
from closed coronal structures decreases as the wind speed increases, thus 
decreasing the spread of FIP bias values measured in the solar wind. 

Second, the Fe/C FIP bias changes with the solar cycle at all speeds, although 
the amount of change is lower for the faster wind streams. This result confirms 
the finding of McIntosh \etal (2011), which found that the Fe/O ratio, another 
proxy for the FIP effect in the solar wind, also shows some degree of variability 
in the fast wind during the decay phase of the solar cycle from 2005 to 2009. 
Further remote sensing observations of the sun-as-a-star by Brooks \etal (2017, 
2018) also indicate that during the rising phase of the solar cycle (2010-2014) 
the FIP bias changes with time: their measurements mostly reflect the composition 
of the denser quiet Sun and active region plasmas where the reconnection-based
slow wind likely comes from. Furthermore, despite the uncertainties in its 
measurement, the abundance of Ne seems to increase during solar minimum relative 
to all other elements, and the amount of increase seems to be lower at larger 
speeds. The present results for the Ne/O ratio confirm those already found by 
Shearer \etal (2014), and discussed by Landi \& Testa (2015) considering also 
remote sensing observations. 

It is interesting to note that Ne has the largest FIP value of all elements 
(21.6~eV), followed by Oxygen and Hydrogen (13.6~eV), Carbon (11.3~eV) and 
Iron (7.9~eV). The present results seem to indicate that solar cycle-induced
FIP effect changes are visible in abundance ratios among elements with large 
FIP differences, such as Fe/C or Ne/X (X being all other heavy elements), and
not in ratios between elements with a similar FIP value (such as C/O).

The variability of the low-FIP to high-FIP ratios in both the in-situ and
remote sensing observations indicate that FIP fractionation processes indeed
somehow change their effectiveness during the solar cycle. The increase in 
the Ne/O ratio confirms this result, as Ne is predicted to be depleted 
relative to all other elements with lower FIP (including the high FIP C, 
N, O and H).

Thus, the present results show that the solar cycle affects solar wind
composition in two main ways: 

\begin{enumerate}
\item By changing the relative abundances of elements with large FIP differences, 
indicating a changing effectiveness of FIP fractionation processes in all types 
of wind and thus in all of their source regions; and 
\item By varying the relative importance of open versus closed magnetic field 
structures in the mixture of source regions for the solar wind slower than 500~km/s.
\end{enumerate}

\subsection{Heating and acceleration processes: ion temperatures and 
differential velocities}

Preferential heating and acceleration of heavy ions over protons have been 
long observed in all types of solar wind, both in situ and close to the Sun.

Remote sensing observations from UVCS (Kohl \etal 1995) have shown that 
O$^{6+}$ flows faster than Hydrogen, and exhibits a remarkably anisotropic 
temperature distribution where the perpendicular temperature is much larger 
than the neutral Hydrogen one (Kohl \etal 2006). These features start to
be detectable in the inner solar corona, beyond 2 solar radii. 

Preferential heavy ion heating has also long been identified in in-situ 
observations of the solar wind, with $T_{ion}/T_p$ ratios found to be 
approximately proportional to mass except in the densest streams (Bochsler 
2007 and references therein). Berger \etal (2011) systematically studied 
ACE/SWICS measurements of heavy ion-proton differential speeds finding 
that most ions travel within $\pm 0.15~C_A$ (where $C_A$ is the Alfven 
speed) of the $\alpha$ speed, which in turn is significantly faster than 
the proton speed, confirming measurements of differential speeds found 
in less systematic measurements by a number of earlier authors. Duruvcova 
\etal (2017) further noted that 1) Helios and WIND measurements of 
$\alpha$-proton differential speeds are larger in faster winds; they 
decrease with distance from the Sun, a trend also found by Reisenfeld 
\etal (2001) to occur in Ulysses measurements up to more than 4~AU, and 
are qualitatively the same at solar minimum and solar maximum.

The present work adds a new temporal dimension to our determination of ion 
heating and acceleration. In fact, our results show that ion and proton 
temperatures follow three interesting trends: 

\begin{enumerate}

\item Their absolute values seem to decrease during solar minimum, although 
the amount of decrease depends on speed itself;
\item The rate of increase of ion temperatures with speed also changes in 
time, being smaller during solar minimum as shown in Figure~\ref{tp_vs_vel}. 
These two results indicate that the ion temperatures of the fast wind
decrease more than those in the slow wind during solar minimum.
\item The dependence on the ion temperatures on the Z/A ratio, and of the
proton-ion temperature ratios on mass, on the contrary, is remarkably 
constant with time.

\end{enumerate}

These results apply both to the general dataset and to the collisionally 
young one, although the slope of the $T_{ion}$ vs speed linear dependence 
is lower than for the full dataset due the higher $T_{ion}$ values of 
collisionally young wind at slow speeds. Also, once the effects of 
Coulomb collisions are minimized, all winds at any speed show the same
dependence on Z/A.

The differential velocity results are much more ambiguous due to the paucity
of data, which forced us to lose velocity resolution, and still have a limited
number of data points. Also, it was not possible to determine whether collisionally
young plasma streams from the slow wind have the same differential speed as those
in the fast wind, as suggested by Stakhiv \etal (2016). Still, we could determine
that the solar cycle did not affect much the differential speed between Hydrogen 
and all other elements, confirming the results of Duruvcova \etal 2017 and 
extending them to all heavy ions.

The origin of ion preferential heating and acceleration has been long studied
(see the reviews of Marsch 2006 and Verscharen \etal 2019). Both phenomena have 
been linked to the action of waves as two sides of the same coin. For example, 
Isenberg \& Hollweg (1982) linked preferential ion acceleration to the same 
Alfven wave dissipation term also responsible for ion preferential heating; 
further refinements involving Alfven wave ion cyclotron resonant heating were 
later discussed by Hollweg \& Isenberg (2002). Other proposed avenues for ion 
heating are high-frequency Alfven wave turbulence (Coleman 1968), and more 
recently stochastic heating by low-frequency Alfven wave turbulence (Chandran 
2010, Chandran \etal 2010, 2013). In all cases, magnetic field fluctuations 
play a central role in the preferential heating and acceleration of coronal 
ions into the solar wind over a range of distances from the Sun.

The present results indicate that the effectiveness of these processes does
change with time. However, results are quite nuanced. The decrease in the 
absolute value of the ion temperatures at all speeds indicates that perpendicular
heating has been weaker during the minimum phase of cycle 24, pointing towards 
a weakening of the heating processes both for protons and for heavier ions at 
all speeds; also, the decrease of effectiveness has been more marked in the 
fast wind than in the slow wind. 

On the other hand, the nature of the process causing preferential heating and
acceleration of heavy ions relative to protons seems to have remained the same.
In fact, not only is there no clear hint at a decrease of differential ion-proton
velocity, but the dependence of the ion temperature values on Z/A, and the 
mass proportionality of ion-to-proton temperature ratios are unchanged for 
all ions at any speed over the entire 1998-2011 period covered by the present 
study.

If wave-particle interactions are responsible for both proton and ion heating, 
and for preferential ion heating and acceleration, our results seem to point
towards a weakening of these interactions. This may be due to a decrease
during the last solar minimum of the wave energy available to heat and
accelerate the solar wind, which would track the decrease in magnetic field 
strength both in the solar wind (Zerbo \& Richardson 2015) and in the 
photosphere (Ingale \etal 2019 and references therein). Still, there are 
no remote sensing confirmation of such a decrease in the solar corona where 
heating and acceleration take place. Such a variation could in principle be 
observable through cycle-related changes in the width of spectral lines 
observed in the solar atmosphere, although such measurements are difficult 
to make due to the lack of systematic spectroscopic observations, to 
line-of-sight effects, and to the presence of two distinct contributions 
(ion thermal and non-thermal speeds) into one single observable (spectral 
line widths). Past attempts at determining cycle-related variations of 
non-thermal motions (Harra \etal 2015), as well as ion temperatures 
(Landi 2007) have found no change, although they were carried out only 
over limited portions of the solar cycle.

\section{Conclusions}
\label{conclusions}

The solar wind is intrinsically a cycle-dependent phenomenon. In the present work,
we have extended this conclusion to the signatures of electron-ion interactions
(wind ionization), elemental abundances, and wave-based heating and acceleration 
(ion temperatures, differential velocity); where possible, we tried to minimize 
the effect of Coulomb collisions on the latter signatures, and investigate whether 
more pristine wind properties change with the solar cycle. We also further refined 
the velocity resolution of our measurements over previous studies. The present work 
clearly shows that all wind properties do change with the solar cycle; specifically

\begin{enumerate}

\item Both proton density and temperature show a significant time dependence,
but while temperature variations clearly follow the 11-year solar cycle, proton 
density shows a steady decline spanning longer timescales with no apparent 
periodic behavior;

\item Both absolute and relative elemental abundances change as a function of the
solar cycle, with the effects being more pronounced in the slow wind than in the 
fast wind;

\item Ion temperatures decrease their absolute value during the 2008 solar 
minimum and begin to recover afterwards; their dependence on wind speed is
affected, but their dependence on the Z/A ratio and on ion mass does not 
change with time;

\item Proton-ion differential speeds are different between the wind slower than 
400~\kms (being almost zero) and the wind faster than 550~\kms (where they range 
between 20-70~/kms), but they do not change with time;

\item Charge state distribution and charge state ratios (including those of Fe)
are strongly dependent on the cycle phase, being lower during solar minimum;

\item Collisionally young wind shows higher ion temperatures at low speed, but
its dependence on mass, Z/A ratio, wind speed and time along the solar cycle does
not change.

\end{enumerate}

While the decrease of proton density seems to act on longer timescales than the
solar cycle, the other plasma properties we investigated do show a clear dependence
on the 11-year solar cycle. Our results point towards the scenario proposed by 
Stakhiv \etal (2016), where both the slow wind and the fast wind are accelerated 
by magnetic waves, but originate from different structures, whose relative importance 
decreases as wind speed increases: the fast wind is present at all speeds, originating 
from open field structures, while the slow wind comes from the reconnection-based 
opening of closed loops and is absent at speeds larger than $\approx$500~\kms. Our 
work adds to this scenario three new features:

\begin{enumerate}

\item The relative importance of open field versus closed field sources in the slow
wind changes with the solar cycle, with the latter greatly decreasing in importance
during solar minimum;

\item The signatures of wave-based heating and acceleration, shown by proton and ion
temperatures, and differential velocities, are present throughout the solar cycle
and indicate that such a process is always active;

\item The effectiveness of magnetic waves at heating and accelerating the solar
wind, as well as shaping its elemental composition, is lower during solar minimum.

\end{enumerate}

\noindent
There are several open questions that still need to be answered. The origin of the
long-term decrease of the proton density is not clear, and also the large variations
in the solar wind charge state distribution can not be explained neither by a decrease 
of electron density and temperature in the solar wind in the inner corona, nor by the
cycle-related variations of EUV and X-ray wind photoionization. Furthermore, ACE/SWICS 
could only sample the solar wind at the ecliptic, whose fast component seldom exceeds 
speeds of 700~\kms: on the contrary, Ulysses showed that out of the ecliptic the wind 
is faster, with a speed that clusters around 750~\kms. Most of this wind, as well as
the solar cycle dependence of its properties still remain to be studied. Also, more 
details are needed in our understanding of the dependence of differential speed on 
both speed and cycle phase, due to the paucity of the data sample we have utilized. 
Hopefully the instruments on board Solar Orbiter (both in-situ and remote sensing) 
will help in answering some of these questions.


\begin{acknowledgements}

E. Landi was supported by NASA grants NNX16AH01G, NNX17AD37G and 80NSSC18K0645. 
The authors would like to thank Prof. S.T.Lepri for assistance and advice on 
the use of ACE/SWICS data.

\end{acknowledgements}


\newpage

\section{References}

\begin{enumerate}

\item Berger, L., Wimmer-Schweingruber, R.F., \& Gloeckler, G. 2011, Phys. Rev. Lett., 106, 151103
\item Berteaux, J.L., Kyr\"ol\"a, E., Quemerais, E., \etal 1995, Sol. Phys., 162, 403
\item Bochsler, P. 2007, Astr. Astroph. Rev., 14, 1
\item Brooks, D.H., Baker, D., van Driel-Gesztelyi, L., \& Warren, H.P. 2017, Nature Comm., 8, 183 
\item Brooks, D.H., Baker, D., van Driel-Gesztelyi, L., \& Warren, H.P. 2018, ApJ, 861, 42
\item Cane, H.V., \& Richardson, I.G. 2003, JGR, 108, 1156
\item Chandran, B.D.G. 2010, ApJ, 720, 548
\item Chandran, B.D.G., Li, B., Rogers, B., \etal 2010, ApJ, 720, 503
\item Chandran, B.D.G., Verscharen, D., Quataert, E. \etal, 2013, ApJ, 776, 45
\item Coleman, P.J. Jr 1968, ApJ, 153, 371
\item Cranmer, S.R., Field, G.B., \& Kohl, J.L. 1999, ApJ, 518, 937
\item Cranmer, S.R., Panasyuk, A.V., \& Kohl, J.L. 2008, ApJ, 678, 1480
\item Cranmer, S.R. 2014, ApJ 791, L31
\item Duruvcova, T., Safrankova, J., Nemecek, Z., \& Richardson, J.D. 2017, ApJ, 850, 164
\item Finley, A.J., Matt, S.P., \& See, V. 2018, ApJ, 864, 125
\item Fludra, A. 2015, ApJ, 803, 66
\item Gloeckler, G., Cain, J., Ipavich, F.M., \etal 1998, Sp. Sci. Rev., 86, 497
\item Harra, L., Baker, D., Edwards, S.J., \etal 2015, Sol. Phys., 290, 3203
\item Heber, V.S., Baur, H., Bochsler, P., \etal (2012), ApH, 759, 121
\item Hollweg, J.V., \& Isenberg, P.A. 2002, JGR, 107, 1147
\item Ingale, M., Janardhan, P., \& Bisoi, S.K. 2019, JGR, 124, 6363
\item Isenberg, P.A., \& Hollweg, J.V. 1982, JGR, 87, 5023
\item Kasper, J.C., Stevens, M.L., Lazarus, A.J., Steinberg, J.T., \& Ogilvie, K.W. 2007, ApJ, 660, 901
\item Kasper, J.C., Stevens, M.L., Korreck, K.E., \etal 2012, ApJ, 745, 162
\item Kasper, J.C., Bennett, M.A., Stevens, M.L., \& Zaslavsky, A. 2013, Phys. Rev. Lett., 110, 091102
\item Khabibrakhmanov, I.K. \& Mullan, D.J. 1994, ApJ, 430, 814
\item Kohl, J.L., Esser, R., Gardner, L.D., \etal 1995, Sol. Phys., 162, 313
\item Kohl, J.L., Noci, G., Cranmer, S.R., \& Raymond, J.C. 2006, Astr. Astroph. Rev., 13, 31
\item Laming, M.J. 2015, Living Reviews Solar Physics, 12, 2
\item Lamy, P., Floyd, O., Quemerais, E., Boclet, B., \& Ferron, S. 2017, J. Geoph. Res., 122, 50
\item Landi, E. 2007, ApJ, 663, 1363
\item Landi, E., \& Testa, P. 2015, ApJ, 800, 110
\item Landi, E., \& Testa, P. 2014, ApJ, 787, 33
\item Landi, E., \& Lepri, S.T. 2015, ApJ, 812, 28
\item Landi, E., Alexander, R.L., Gruesbeck, J.R., \etal 2012, ApJ, 744, 100
\item Landi, E. \& Cranmer, S.R. 2009, ApJ, 691, 794
\item Lepri, S.T., Landi, E., \& Zurbuchen, T.H. 2013, ApJ, 768, 94
\item Luhmann, J.G., Li, Y., Arge, C.N., Gazis, P.R., \& Ulrich, R. 2002, JGR, 107, 1154
\item Marsch, E. 2006, LRSP, 3, 1
\item McComas, D.J., Ebert, R.W., Elliott, H.A., \etal 1998, G. Geoph. Res., 35, L18103
\item McComas, D.J., Angold, N., Elliott, H.A., \etal 2013, ApJ, 779, 2
\item McIntosh, S.W., Kiefer, K.K., Leamon, R.J., Kasper, J.C., \& Stevens, M.L. 2011, ApJL, 740, L23
\item Owens, M.J., Cliver, E., McCracken, K.G. \etal 2016, J. Geoph. Res., 121, 6048
\item Reisenfeld, D.B., Gary, S.P., Gosling, J.T., \etal 2001, JGR, 106, 5693
\item Richardson, I.G., \& Cane, H.V. 2010, Sol. Phys., 264, 189
\item Shearer, P., von Steiger, R., Raines, J.M., \etal 2014, ApJ, 789, 60
\item Smith, C.W., L'Hereux, J., Ness, N.F., \etal 1998, Sp. Sci. Rev., 86, 613
\item Tracy, P.J., Kasper, J.C., Zurbuchen, T.H., \etal 2015, ApJ, 812, 170
\item Tracy, P.J., Kasper, J.C., Raines, J.M., \etal 2016, Phys. Rev. Lett., 116, 255101
\item Verscharen, D., Klein, K.G., \& Maruca, B.A. 2019, LRSP, 16, 5
\item Yamauchi, Y., Suess, S.T., Steinberg, J.T., \& Sakurai,  T. 2004, JGR, 109, A03104
\item Yermolaev, Y.I., Lodkina, I.G., Khokhlachev, A.A., \etal 2021, JGR, 126, e2021JA029618
\item Wang, Y.-M., 1998, ASP Conf. Series, 154, 131
\item Zerbo, J.-L. \& Richardson, J.D. 2015, J. Geoph. Res., 120, 10250

\end{enumerate}

\newpage

\end{document}